\def\year{2021}\relax
\pdfoutput=1

\documentclass[letterpaper]{article} 
\usepackage{aaai21}  
\usepackage{times}  
\usepackage{helvet} 
\usepackage{courier}  
\usepackage[hyphens]{url}  
\usepackage{graphicx} 
\usepackage{natbib}  
\usepackage{caption} 
\usepackage{mathtools}
\usepackage{subfigure}
\usepackage{moreverb,url}
\usepackage[bookmarksopen,bookmarksnumbered]{hyperref}
\usepackage{url}

\PassOptionsToPackage{hyphens}{url}
\newcommand\BibTeX{{\rmfamily B\kern-.05em \textsc{i\kern-.025em b}\kern-.08em
T\kern-.1667em\lower.7ex\hbox{E}\kern-.125emX}}

\title{Beyond Algorithmic Bias: A Socio-Computational Interrogation of the Google Search by Image Algorithm}

\author {
    Orestis Papakyriakopoulos,\textsuperscript{\rm 1}
    Arwa Michelle Mboya\textsuperscript{\rm 2} \\
}
\affiliations {
    \textsuperscript{\rm 1} Center for Information Technology Policy, Princeton University\\
    \textsuperscript{\rm 2} Center for Civc Media,
    Massachusetts Institute of Technology\\
    orestis@princeton.edu, michelle.mboya1@gmail.com
}

\begin{document}
\maketitle

\begin{abstract}

We perform a socio-computational interrogation of the google search by image algorithm, a main component of the google search engine. We audit the algorithm by presenting it with more than 40 thousands faces of all ages and more than four races and collecting and analyzing the assigned labels with the appropriate statistical tools. We find that the algorithm reproduces white male patriarchal structures, often simplifying, stereotyping and discriminating females and non-white individuals, while providing more diverse and positive descriptions of white men. By drawing from Bourdieu's theory of cultural reproduction, we link these results to the attitudes of the algorithm's designers, owners, and the dataset the algorithm was trained on. We further underpin the problematic nature of the algorithm by using the ethnographic practice of \textit{studying-up:} We show how the algorithm places individuals at the top of the tech industry
within the socio-cultural reality that they shaped, many times creating biased representations of them. We claim that the use of social-theoretic frameworks such as the above are able to contribute to improved algorithmic accountability, algorithmic impact assessment and provide additional and more critical depth in algorithmic bias and auditing studies. Based on the analysis, we discuss the scientific and design implications and provide suggestions for alternative ways to design just socioalgorithmic systems.

\end{abstract}




\section{Introduction}

Scientists and philosophers have long stated that any type of technological medium de facto transforms the social, cultural, and political relations existing in that world \cite{winner1980artifacts, mcluhan1967medium}. Algorithms are no exception, as their invasion and ubiquity in every aspect of society constantly redefines human sociability, political conduct and social structures. Given this, scientists investigate how algorithms impact society \cite{pasquale2015black}, the context of their use \cite{selbst2019fairness}, the conditions of their creation \cite{karkkainen2019fairface} and properties of their structure \cite{yang2020towards}. These investigations can take place on a theoretical level \cite{blodgett2020language} or a mathematical level \cite{barocas_hardt_narayanan}, can take place by testing already existing algorithmic implementations in the industry and society \cite{buolamwini2018gender}, by evaluating the people and social groups who are influenced by the algorithms \cite{woodruff2018qualitative}, or by evaluating the algorithms' designers themselves \cite{barrett2019platform}. 

Furthermore, scientists develop frameworks that can assist researchers and policymakers with the systematic analysis, evaluation, design and governance of algorithms \cite{mittelstadt2016ethics}.  Selbst et al. \shortcite{selbst2019fairness} stated that the use of social theories (e.g. from science and technology studies) can contribute to the development of context-aware machine learning applications. In the case of using computation as means for locating algorithmic injustice, Abebe et al. \shortcite{abebe2020roles} described specific roles that computational analysis can take, in order to be able to confront deeper patterns of inequality, which many studies related to algorithmic bias fail to define and take into consideration \cite{blodgett2020language}. Barabas et al. \shortcite{barabas2020studying} argued that there is also a need to "study up". That is, to reorient algorithmic studies, by focusing on the relationship between socially dominant groups and technological artifacts. This can contribute to further knowledge generation about algorithms, as well as provide critical perspectives of society in algorithmic studies.

\textbf{Statement of the problem}

By taking into consideration the proposals to rethink the use of computational tools,  the need to complement them with social theoretic work, and the call to reorient algorithmic studies upwards,  this study investigates the google search by image algorithm, a basic component of the google search engine. Also known as reverse image search, the algorithm takes a picture as input, and returns a label for that image, related links, as well as visually similar images  (figure \ref{tribe_african}a). Similar to prior work that has shown that various components of the google search engine might result in racist inferences, discriminate against social groups, and (re)produce social power and information asymmetries \cite{noble2018algorithms, vincent_2018, Diaz2008, urman2021matter}, the same applies for the google search by image algorithm. For example, as figure \ref{tribe_african}b shows, the algorithm labels a picture of an African woman in the traditional tribal uniform as "most black women in the world." Provoked by cases such as this, we uncover how the google search by image algorithm (re)produces social structures and biases by investigating how it views people of different races, genders, and ages. To draw awareness about the algorithm's function and by "studying up", we evaluate how the algorithm views key-stakeholders in the tech industry, and show how these actors are placed in the sociopolitical reality that they have designed. To achieve the above, we perform a socio-computational interrogation and seek the answer to the following research questions:

\begin{figure}[t]

\centering
\fbox{\begin{minipage}{3.25in}
(a)

\begin{subfigure}
  \centering
   \includegraphics[width=1\linewidth]{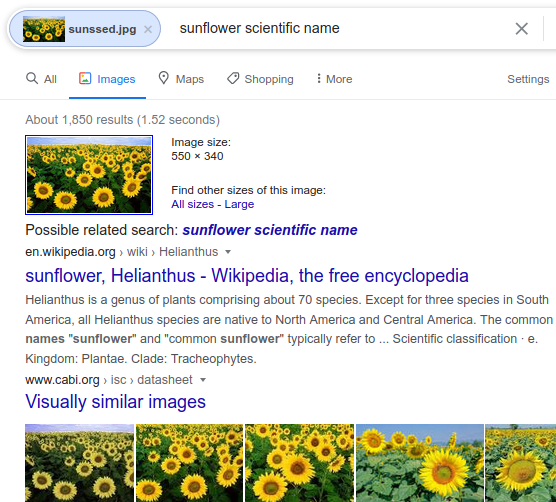}
\end{subfigure}%
(b)

\begin{subfigure}
  \centering
   \includegraphics[width=1\linewidth]{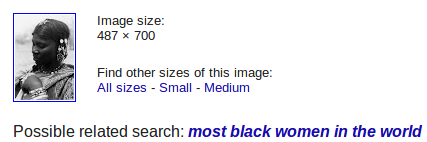}
\end{subfigure}
\end{minipage}}
\caption{(a) Overview of the google search by image algorithm. For a given image the algorithm generates a label, and returns similar images and potentially related URLs. (b) Label returned by the algorithm when shown an African woman in tribal dressing.}

\label{tribe_african}
\end{figure}

\begin{description}

\item[RQ1:]\textbf{How does the google search by image algorithm see people of different races, genders, and ages?}

\item[RQ2:]\textbf{Which cultural and social structures are (re)produced by the algorithm?}

\item[RQ3:]\textbf{What happens when individuals on the top of the techno-hierarchical ladder become the objects of the algorithm?}
\end{description}

\textbf{Significance of the study \& contributions}

The study provides novel insights into an understudied component of the most popular search engine in the world. Not only does it show how the algorithm views people of different genders, races, and ages, but also evaluates generated knowledge, uncovering sociopolitical structures immanent in the algorithm. The study also critically confronts individuals who have the power to change algorithmic inferences, by placing them in the algorithm's sociopolitical reality. The study serves as a scientific and design provocation and provides following concrete contributions: 

\begin{itemize}

\item We evaluate the google search by image algorithm's culture by drawing from Bourdieu's theory of cultural reproduction, treating it as a subject and interrogating it through computational means. The interrogation takes place by showing the algorithm more than 48 thousand images of people of different ages, genders, and races (prompts) and analyzing the collected labels (responses) with quantitative and qualitative tools. Since we want to give minimal stimuli to the algorithm to uncover its social perceptions, the images are people's portraits and we specifically investigate how the algorithm evaluates human appearance.

\item We find that the algorithm's responses focus on following categories: celebrity status, gender, age,  external features (beauty, shape, hairstyle), sentiment, ethnic and racial heritage and location.

\item We show that label selection is dependent on an individual's race and gender, both in terms of the language in which the algorithm will use and the content of the label. 

\item We find that the algorithm (re)produces sociopolitical and cultural structures that exist in the white patriarchal society. This is visible in the algorithm's simplified description of women and non-white individuals, its frequent stereotyping, and its inability to make accurate predictions about people. Furthermore, the algorithm exoticizes non-white individuals in terms of beauty, associates them with negative sentiments, and tends to assign more adjectives related to their physique and external features. In contrary, the algorithm relates white males more often with terms about character virtues and higher social status, as well as provides richer descriptions about them.

\item We study up individuals that are in high positions within the techno-hierarchical ladder. By showing images of them to the algorithm, we illustrate how the algorithm places them in the already described culture, creating biased representations of them. In this way, we provoke them to reflect on the influence algorithms such as this have on people, and to consider ways to bend technology towards justice.

\item Based on the analysis, we discuss the benefits of integrating social-theoretic frameworks such as the above in algorithmic bias and auditing studies. We further comment on the scientific and design implications, point out the ethical issues we faced in performing our investigation, and propose scientific and design alternatives that can work towards socio-algorithmic justice.
\end{itemize}

\section{Theoretical approach }
\subsection{Beyond algorithmic bias}

When an algorithm becomes the object of scientific investigation, the tactics to investigate it are also defined by the limits and possibilities of the extracted knowledge. To understand an algorithm's culture, we combine theoretic and computational means. We do so because pure algorithmic bias studies often lack the ability to associate algorithms with important cultural and sociopolitical aspects \cite{abebe2020roles,blodgett2020language}. On the other hand, theories coming from philosophy, sociology, science and technology studies, and ethnography, have successfully generated knowledge about cultural and political aspects of algorithms \cite{seaver2017algorithms, cave2018ai}, as well as unfair and problematic outcomes \cite{mittelstadt2016ethics,selbst2019fairness}. Therefore, we argue that quantitative studies, be those algorithmic bias analyses or algorithm audits, can benefit from incorporating frameworks coming from outside fields such as the above. 

We locate the benefits of combining theoretic and computational means for knowledge extraction in four distinct dimensions. First, studying algorithms based on social theories helps us understand the algorithms and the inferences they produce. Even studies that investigate algorithmic inferences purely mathematically \cite{obermeyer2019dissecting, papakyriakopoulos2020bias} conclude that it is inadequate to analyze them apart from the social conditions in which they were created, be that an NLP architecture or an automated decision making system. Second, an algorithm's placement in a social reality leads us to a better understanding of the world. For example, combining and interpreting social and algorithmic patterns dissolves the false dichotomy of the digital and the real and gives a holistic picture of the world we are all situated in  \cite{haines2017towards, murray2019entangled}. It also provides knowledge on how social groups that are the objects of algorithmic investigations are positioned in society, as well as which general structures of power and ethics are being reproduced. Third, such an analysis of algorithms uncovers features of the coupling between algorithms and society. It detects how algorithms prescribe objectivity, learns which social groups are in- and which excluded, and shows how an algorithm can lead to \textit{calculated publics} \cite{gillespie2014relevance}, that is, how the objects of algorithmic inferences learn to perceive themselves. Furthermore, the social context-aware analysis of an algorithm leads to an understanding of how an algorithm participates in a social reality, how it composes the social reality, and how it becomes the social reality \cite{neyland2019everyday,van2019hiring}, all the while transforming what is considered ethical, acceptable and normal. Last but not least, a socio-computational study of algorithms provides the benefit of computation: the deployment of algorithms for generating knowledge about a part of the world provides researchers a vast amount of structured information that can complement and support any qualitative analysis \cite{christin2020ethnographer,markham2016ethnography, abebe2020roles}. When algorithms become the object of investigation, computation can function as an especially ideal mediator between researchers and algorithms, acting as a translator of thought between the investigator and the investigated.

\subsection{Studying Up Technology}

Any knowledge extracted during the practice of developing, deploying, or analyzing an algorithm is \textit{situated knowledge} \cite{elish2018situating,haraway1988situated}, that is knowledge that is dependent on the social and political context adopted by each individual when interacting with the algorithm (consciously or unconsciously). This perspective dependent knowledge generation process allows researchers to elect the appropriate framework to investigate algorithms and answer their descriptive or prescriptive research questions. Building on this, Barabas et al. \shortcite{barabas2020studying} called for researchers to \textit{study up} in the field of algorithmic fairness. \textit{Studying up} \cite{nader1972up} denotes the reflexive analysis of the upper end of the social power structure when investigating a cultural system. In the case of algorithms, studying up includes the focus of the analysis on the powerful actors of an algorithmic case study, be that tech companies, regulators, or social groups and individuals that benefit the most from the existence of algorithmic systems. As a practice, studying up provides three concrete benefits. First, it is a more integrative scientific practice, because it allows knowledge generation on a generally understudied part of the social hierarchy. Second, it promotes a democratized scientific paradigm, since power asymmetries become the epicenter of an investigation. Third, it allows researchers to investigate new objects that might be more interesting to them, avoiding the re-analysis of objects and structures that have been studied repeatedly in the past. 

We perform the socio-computational interrogation of the google by image algorithm also by studying powerful individuals in the tech industry in order to bring new perspectives and provocations in the community of algorithmic fairness, which holds a peculiar social structure in itself. Specifically, D' Ignacio and Klein \cite{d2020data} argue in their work that a large part of data ethics research is shaped and promoted by the very institutions that have yet to become the objects of critical scientific investigation. This can have hindering scientific effects, since the boundary between studier and studied are blurred, raising some troubling ethical and political concerns, which technoscience has sought to overcome for a long time  \cite{forsythe1999ethics}. To constructively contribute to the debate, we adopt a relational thinking framework when performing the interrogation \cite{stich2015thinking}. We exploit the nature of situated knowledge and we reflect on what it means for an individual to be on the top of the tech industry as an algorithm designer, but also how their privilege is obscured when they become objects of the algorithmic inferences. We also adopt this relational thinking in studying up to ourselves. Since the study is performed by a black female and a white male, we reflect on what this means when we interpret the specific algorithmic inferences in emic terms, and how we come to the final intersubjectively generated knowledge \cite{peters2016up}. Furthermore, we acknowledge our individual position in the social hierarchy, and the automatic limitations this poses in holistically uncovering all of its features.

\section{Conceptual framework \& interrogation tactics}

An ideal way to sociopolitically evaluate the google search by image algorithm is to interrogate, investigate, and understand how the algorithm's designers, engineers, and owners conceptualized and shaped the artifact. This can explain how practices, social context, histories and values shape the algorithm's culture \cite{denton2020bringing,seaver2018should}. Nevertheless, the sociotechnical systems of proprietary algorithms such as the one we are analyzing are extremely opaque, making it infeasible to perform an analysis of that magnitude and type \cite{burrell2016machine,alvarado2017big}. To overcome this issue, we adopt an alternate framework that follows Bourdieu's theory of cultural reproduction  \shortcite{bourdieu1990reproduction},  treating the algorithm itself as a subject that was shaped by a specific culture, and as any individual, it (re)produces the conditions of its creation. The theory dictates that an individual's world views, behaviours, and language will contain and replicate the values, culture, and power structures of the society they were created in. As a child being educated in a society replicates the values and behaviors taught at school and the family, the algorithm will also replicate values based on its input data, the decisions of the machine learning practitioners that created it, as well as the company that it owns it. This allows us to overcome issues of opacity, since we can easily interact with the algorithm through the google website and by learning about it we can also learn about the overall socio-technical system which includes the algorithm developers, owners, and designers. In this way, we can obtain an integrative view of the algorithm's culture. We acknowledge that the field of cultural studies holds a critical position against quantification \cite{dixon2016diffractive}. Nevertheless, we claim that by using such a theoretic framework to evaluate quantitative results, we are able to provide additional depth to our analysis. Furthermore, we chose Bourdieu's cultural reproduction framework not only because it suits our study, but because it shows that algorithmic bias studies can benefit from social scientific frameworks in ways that are not directly visible, or were not explicitly developed for understanding technological systems.   

\begin{figure}[t]
  \centering
  \frame{\hspace{0.35cm}\includegraphics[width=0.33\textwidth]{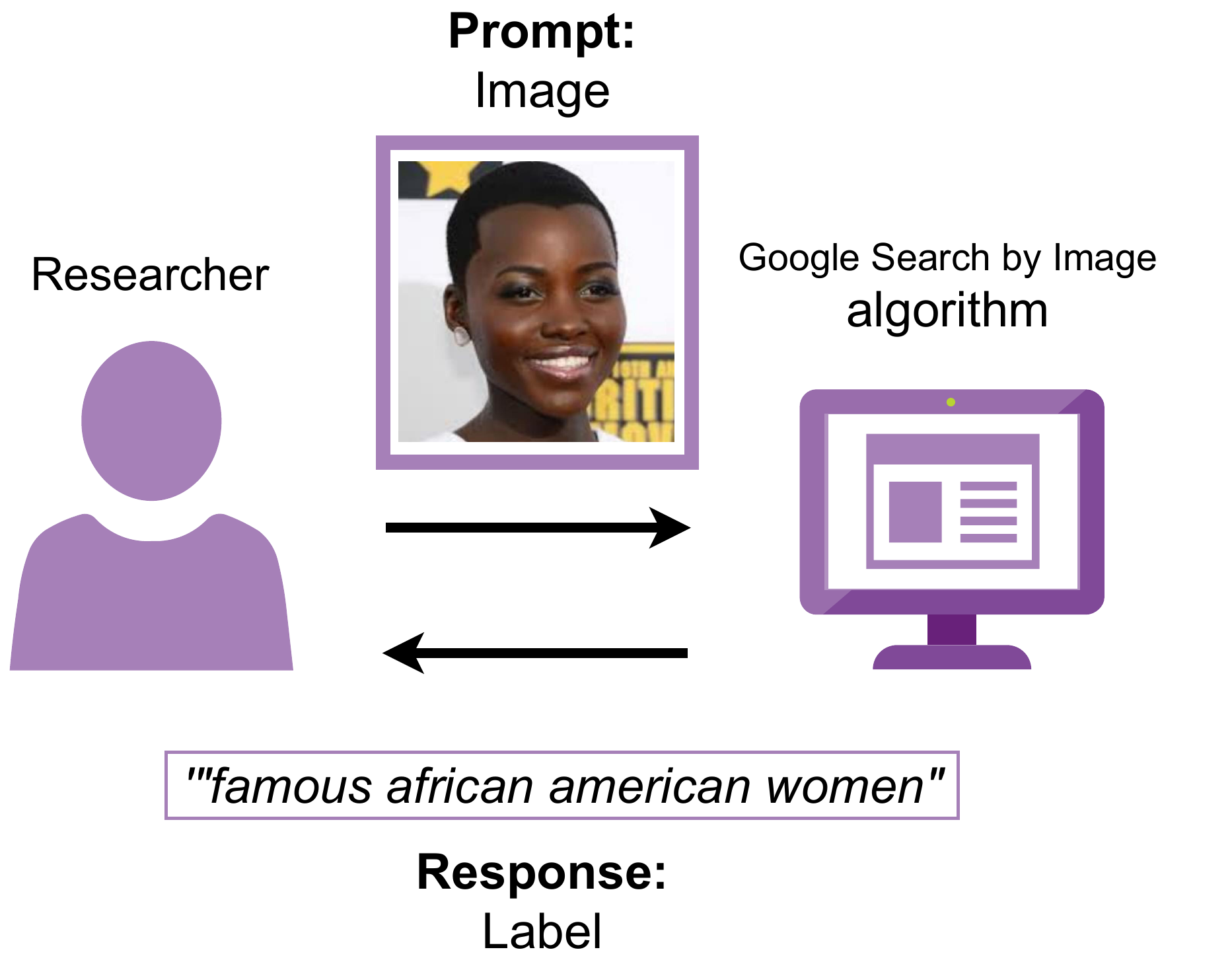}}
  \caption{Overview of the interrogation tactics. The researchers show the algorithm images (prompts) and collect and analyze the corresponding labels (responses).}
  \label{interview}
\end{figure}

To extract information from the algorithm-subject, 
we construct a pipeline, which allows us to systematically and rigorously audit it \cite{sandvig2014auditing} in the form of an interrogation (figure \ref{interview}). Since the algorithm takes images as input, the selection of an image and the feeding of it into the algorithm denotes the process of prompting the subject. The algorithm, in return, provides a label for each image it receives - the response. Nevertheless, each image and the corresponding label provides limited knowledge to us about how the algorithm sees the world. Therefore, we develop a data-intensive process that allows us to show thousands of images to the algorithm and save the generated responses. The process is structured as follows: First, we create datasets of images that comprise the pool of our prompts. Then, we program an automated anonymous browsing agent that crawls the google images website, iteratively uploading each image in the dataset and then saving the algorithm's labels (responses). The anonymity of the browser ensures that the google search engine will personalize their responses to a minimal degree. Afterwards, we employ qualitative techniques and computational tools to understand the collected information about the algorithms' culture through it's handling of people's faces and bodies. The performed socio-computational interrogation has the advantage of scaling \cite{hsu2014digital}, improves transparency and replication \cite{abramson2018promises}, and enhances algorithmic bias research, by incorporating a social scientific perspective when evaluating algorithm's behavior.

The primary scope of the study is to understand how the algorithm views people of different races, genders, and ages (RQ1), and based on that to uncover cultural, institutional, and social structures reproduced by it (RQ2). Furthermore, we study up powerful actors in the tech-industry (RQ3), by investigating how the algorithm perceives them. To perform these tasks, we split the interrogation into two distinct parts: exploratory prompting and focused exploration.  Exploratory prompts are the input pictures we show to the algorithm in order to uncover general regularities in its perceptions. Focused exploration, on the other hand, encompasses additional computational analyses that aim to investigate how the subject organizes knowledge about a specific topic.

\subsection{Exploratory Prompts}

\begin{figure}[t]
  \centering
  \includegraphics[width=0.45\textwidth]{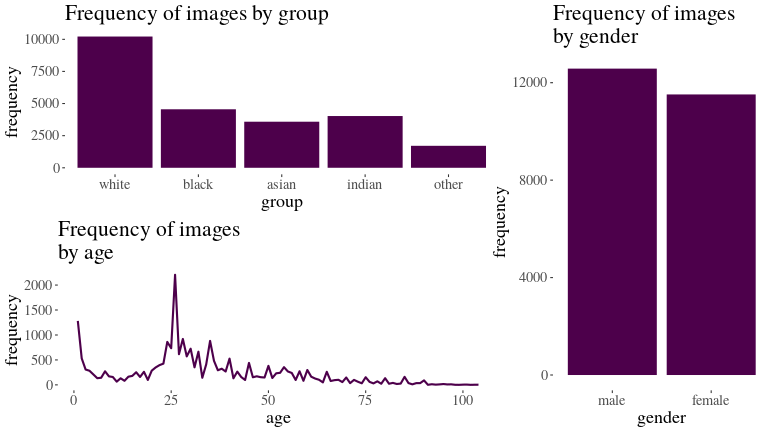}
  \caption{Distribution of metadata categories race, age, gender in the UTKFace dataset}
  \label{dataset}
\end{figure}

To investigate how the algorithm views people of different genders, ages, and races (RQ1) we form our exploratory prompts accordingly. Exploratory prompts consist of 24.100 images coming from the UTKFace dataset \cite{zhifei2017cvpr}. The dataset contains faces of people of at least five races (asian, black, indian, white, and other, which primarily includes Latinx individuals), of age between 0 and 116 years, and of male and female gender. The ratios and frequency of classes in the above categories can be found in figure \ref{dataset}. The original dataset creators collected the images from the internet, and images include people of various poses, facial expressions, and in different backgrounds.

As the dataset creators state, the labels created for each image were generated by the DEX algorithm \cite{rothe2015dex} and double checked by a human coder. Given this, the data generation process of the dataset automatically poses specific restrictions to our study and directly situates the knowledge created. First, since the ground truth values are algorithmic predictions, they are approximations of the actual ones, inserting a bias in our study. Second, race is a social construct, hence the given racial categories are a product of the views of the algorithm's creators, and not of the individuals in the dataset. Third, the algorithm used classified people in binary genders, automatically erasing non-binary genders. We acknowledge that these facts automatically pose a limitation in our analysis and is a serious ethical concern that should be taken into consideration by the whole scientific community. We decided to use the features in our analysis, as they give important insights about the google algorithm, but further discuss the issue later on and included it in the study's scientific and design implications.

\begin{figure}[t]
  \centering
  \includegraphics[width=0.42\textwidth]{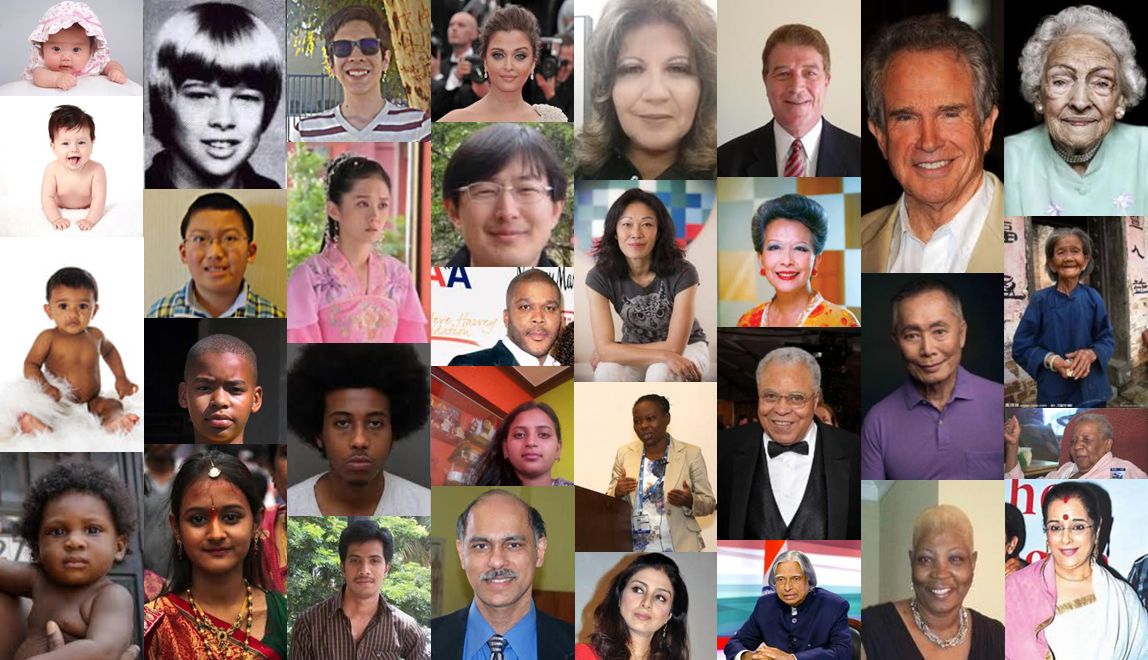}
  \caption{Exemplary images in the UTKFace dataset \cite{zhifei2017cvpr}}
  \label{dataset_examples}
\end{figure}

To gather adequate information about the algorithm's views, we show the images to it in two formats. The first format includes the original photos, which might include the whole body of an individual, and might show the context and background in which the individual is in. The second format is the close-up, with images containing only the face of an individual. In this way, we extract as much information from the algorithm as possible, and see what is of interest to it in its answers. At the exploratory prompts phase, we feed the algorithm a total of 48.200 images (examples in fig. \ref{dataset_examples}). We purposefully run the automated browsing agent by switching to a German IP, in order to investigate whether the algorithm personalizes its answers according to how they perceive us as users. We use the google translate API \cite{google} to detect the languages in which the algorithm returns its responses. We create weighted graphs of the most frequently assigned labels for male and female individuals across racial groups, as well as measure the entropy of labels' distribution for each race-gender subgroup. The equation has the form:  

\begin{align*}
\mathrm {H_{subgroup}} (X)=-\sum _{i=1}^{n}{\mathrm {P} (x_{i})\log \mathrm {P} (x_{i})}
\end{align*}

, where $H$ is the subgroup-specific entropy value, $n$ is the total number of labels appearing for a subgroup and $P(x_i)$ is the probability of appearance for a subgroup-specific label. Mathematical entropy serves as a measure that quantifies how diverse labels are for each population, with higher entropy corresponding to higher diversity. Furthermore, we use the pre-trained language models of spacy \cite{spacy2} and perform named entity recognition and part-of-speech-tagging. In this way, we extract important information about the language and content in the algorithm's inferences. Next, we go through all the labels manually, and extract frequent appearance adjectives assigned to the different social groups and genders, as well as stereotypes and simplifications. We also locate and evaluate unexpected labels that the algorithm returns. We visualize some of these findings through word-clouds, in order to make algorithmic inferences more tangible to the reader. The above results help us understand how the algorithm views society (RQ1) and the cultural and social structures (re)produced by the algorithm (RQ2).

\subsection{Focused exploration}

\begin{table}
\centering
\caption{List of tech industry stakeholders that we study up.}
\resizebox{2.5in}{!}{\begin{tabular}{cccc}
\textbf{Individual}      & \textbf{Company}   & \textbf{Position}  & \textbf{Gender} \\ \hline
Mark Zuckerberg & Facebook  & CEO       & M      \\
Sundar Pichai   & Google    & CEO       & M      \\
Sheryl Sandberg & Facebook  & COO       & F      \\
Jeff Bezos      & Amazon    & CEO       & M      \\
Satya Nadella   & Microsoft & CEO       & M      \\
Peter Thiel     & Palantir  & Founder   & M      \\
Larry Page      & Google    & Founder   & M      \\
Sergey Brin     & Google    & Founder   & M      \\
Ginni Rometty   & IBM       & CEO       & F      \\
Susan Wojcicki  & YouTube   & CEO       & F      \\
Safra Catz      & Oracle    & CEO       & F      \\
Eric Schmidt    & Google    & EX CEO    & M      \\
Ruth Porat      & Alphabet  & CFO       & F      \\
Amy Hood        & Microsoft & CFO       & F      \\
Gwynne Shotwell & SpaceX    & President & F      \\
Marissa Mayer   & Yahoo     & CEO       & F      \\
Tim Cook        & Apple     & CEO       & M      \\
Bozoma St. John & Uber/WME  & CBO       & F     \\ \hline
\end{tabular}}
\label{techtable}

\end{table}

After getting a general overview of the algorithms inferences, we perform a focused exploration to deepen our understanding of the algorithm's culture (RQ1 \& RQ2). Given the appearance of specific regularities between labels and social groups, we develop a multinomial logistic regression model to quantify and confirm these associations, as well as to uncover more.  In the model we use labels as our dependent variable and gender, racial group, and age as independent variables. We create a list of 39 abstract adjectives (labels) that we insert into our model, each of them relating to a person's appearance, character, sentiments, size, social status, or other feature.  We replace any other label aside the above 39 with the $other\_token$ label, which we also use as baseline label in our model. The model has the form: 

\begin{align*}
& Pr(Y=C)=\frac {e^{{\boldsymbol {\beta }}_{C}\cdot \mathbf {X}}}{1+\sum _{k=1}^{K-1}e^{{\boldsymbol {\beta }}_{k}\cdot \mathbf {X}}}
\\ & \text{with} \\
& \beta_{C}\cdot \mathbf {X} = \beta_{0,c} + \beta_{1,c} \cdot age +  \beta_{2,c} \cdot female +  \\
& \beta_{3,c} \cdot black  + \beta_{4,c} \cdot asian + \beta_{5,c} \cdot indian + \beta_{6,c} \cdot other 
\end{align*}

, where $C$ is a label in the dataset, $K$ is the total number of labels,  $\beta_{i,c}$ are label-specific estimators, $age$ is the continuous variable depicting the age, and $black$, $asian$, $indian$, $other$ are indicator variables representing social groups.  In order to continue studying up, we use the white male as the baseline population for our model in order compare how differently the algorithm depicts white men as compared with groups lower in the social hierarchy. We evaluate the results based on the existing sociopolitical and socio-historical conditions in our society, uncovering biases but also showing cases where the algorithm deconstructs white male privilege. We also investigate the algorithms predictive abilities. Prior research shows that commercial algorithms discriminate based on classes such as gender and race, by making poor predictions for those categories \cite{buolamwini2018gender}. Therefore, we compare whether gender-specific labels generated by the algorithm identify individuals' gender in the same way as in the actual metadata in the dataset. We compare the agreement rates for all social groups and investigate whether there is systematic discrimination of individuals of a specific race. Taking into consideration the above results, we apply Bourdieu's theory of cultural reproduction to critically evaluate the traced associations. 

We use the second part of our focused exploration to study up the techno-hierarchical ladder (RQ3). Towards this purpose, we create a dataset of 180 images that contain captions of 18 individuals that are former or current C-suite executives, founders or presidents of major tech companies (9 male and 9 female, table \ref{techtable}). Our rationale is that these individuals hold the actual power to decide whether and how an algorithm such as the one we are studying will be deployed, set the ethical norms that products are designed according to, as well as make decisions that affect billions of people every day. Since tech companies' algorithms classify individuals constantly, the above powerful actors are the actual human classifiers of the world. Nevertheless, as Bourdieu  \shortcite{bourdieu1987makes} states, actors are always simultaneously classifiers and classified. Since these actors are in the highest class of of the socio-algorithmic world, their decisions on how to classify are set by their own status in the hierarchy. Therefore, as a provocation, we want to show how the algorithm classifies them when they are stripped of their social privilege. Under normal conditions, the google search by image classifier directly recognizes these individuals, and returns a label with their name. Nevertheless, when shown to the algorithm images that are not indexed by google, the algorithm fails to recognize them by name. Therefore, we search for YouTube videos containing the actors and created snapshots of the individuals, which comprise the dataset presented above. Then we feed the images to the algorithm and evaluate how the algorithm treats the tech companies' key-stakeholders when they are reduced to ``normal'' humans.

\section{Socio-computational interrogation}

In the following, we present the results of our socio- computational interrogation. We split it into three parts. The first part corresponds to how the algorithm recognized the world generally. The second part illustrates more concrete group perceptions and associations. The third part evaluates the algorithm's culture, and offers a critical provocation of the techno-hierarchical ladder.

\subsection{An algorithmically visioned society}

\begin{table}
\caption{Categorization of terms appearing in algorithm's labels, their frequency, and top examples.}
\resizebox{3.3in}{!}{\begin{tabular}{c|cc}
\textbf{Entity}        & \textbf{ratio} & \textbf{Top terms}                                                                                                                  \\ \hline
Celebrity              & 13.6\%           & Winona Ryder, Jin, Jackie Chan                                                                                                       \\[5pt]
Racial or Ethnic Group & 6.9\%            & Indian, Asian, Black, White, Latino                                                                                                  \\[5pt]
Age                    & 4.3\%            & 60 years old, 10 years old, 13 year old                                                                                              \\[5pt]
Location               & 2.1\%            & India, China, Hollywood                                                                                                              \\[5pt]  \hline  \hline

Noun                   & 45\%             & girl, baby, man, hair, year                                                                                                          \\[5pt]
Adjective                   & 15\%             & beautiful, short, cute, hot, famous                                                                                                  \\[5pt]
Verb              & 5.5\%            & looking, dating, smiling                                       
\\ \hline
\end{tabular}}
\label{terms}

\end{table} 

\begin{figure}[t]
  \centering
  \includegraphics[width=0.45\textwidth]{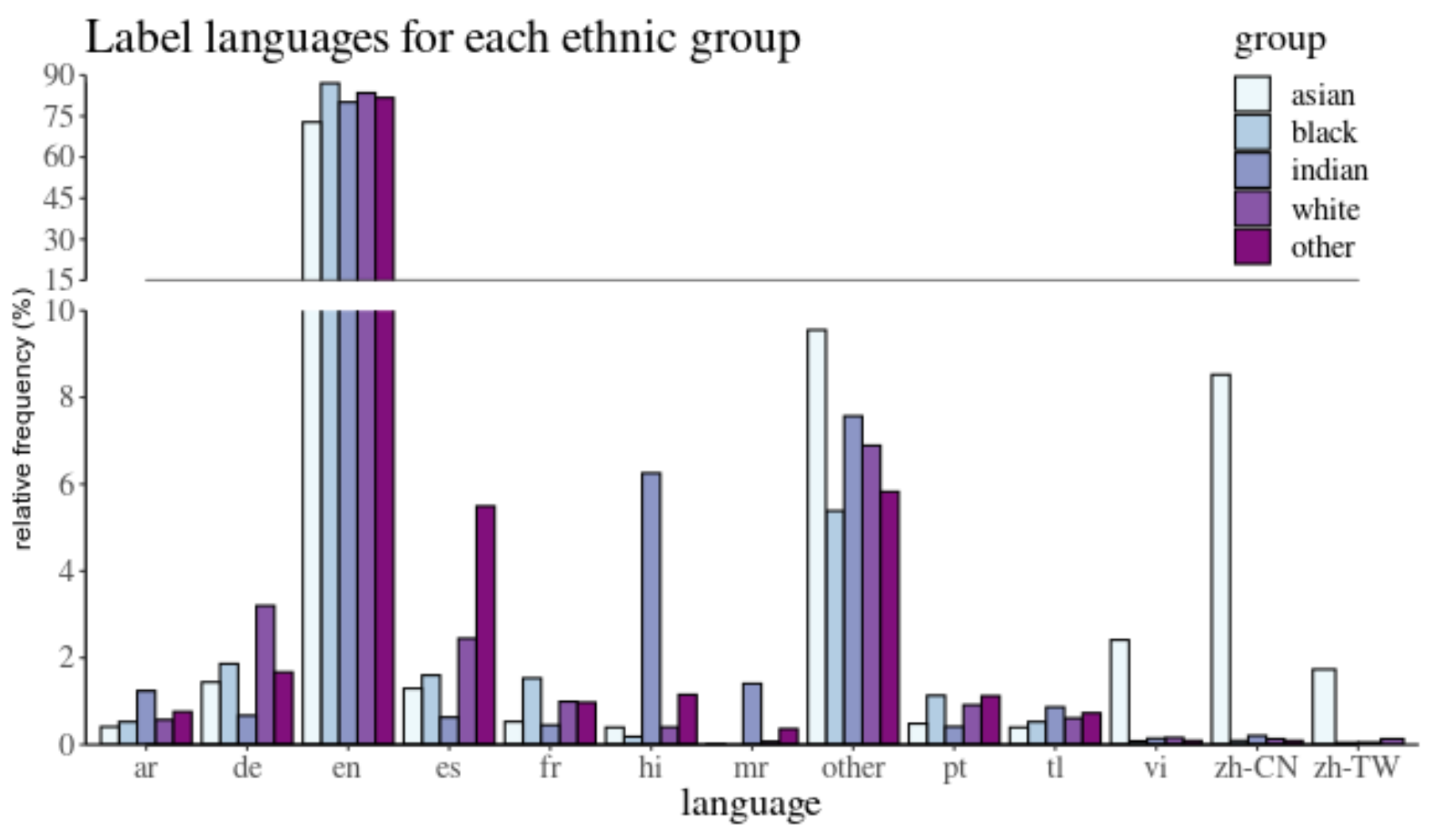}
  \caption{Distribution of the top 13 label languages across different racial categories}
  \label{language}
\end{figure}

The google search by image algorithm gave 17,944 unique answers to 48,200 exploratory prompts. Since we asked the algorithm to label individuals' portraits that were not characterized by specific behaviours, the algorithm's replies focused, as intended, primarily on people's appearance and external characteristics. Some label examples were the following: ``black men hair twist'', ``old lady on beach'', ``Italian actors over 50'',  ``happy african with laptop'', or ``indian sweet girl''. As expected, most terms were appearance-related, describing the gender, age, racial or ethnic origin, attractiveness, shape, sentiments, or hairstyle of individuals. 
Table \ref{terms} provides the main term categories existing in the labels, as well as the frequency of their appearance. Since our dataset comprised of images collected from the internet, a proportion of it included celebrities. Therefore, the most common term category was celebrity names. In most cases, the algorithm recognized them and returned their name as a label (13.6\% of the total labels included the name of a celebrity). The second most common label category was Racial or Ethnic Group, labels that were related to ethnic descriptions inferred by the algorithm (6.9\%). The algorithm also inferred the age of the person, as well as the location of the picture (4.3\% and 2.1 \% of the total images respectively). Overall, the algorithm used nouns for 45\% of the labels, most of which being gender-specific (e.g. girl, gentleman). They also used adjectives for 15\% of the labels (excluding race and ethnicity related ones), and verbs for 5.5\% of the labels. Usually, each label combined more than one of the above categories.

\begin{table}
\caption{Gender classification agreement by race between the initial dataset and google search by image algorithm's predictions.}

\resizebox{3.3in}{!}{\begin{tabular}{lll|ll}
    & \multicolumn{2}{c}{Asian}     & \multicolumn{2}{c}{Black}                                 \\ \hline
                              & Female\textsubscript{pred}    &  Male\textsubscript{pred}      & Female\textsubscript{pred}                   & Male\textsubscript{pred}                 \\
Female\textsubscript{ground}                   & 0.97 & 0.03 & 0.97                 & 0.03             \\
Male\textsubscript{ground}                      & 0.12 & 0.88 & 0.07                & 0.93             \\
\multicolumn{1}{c}{\textbf{}} & \multicolumn{2}{c}{Indian}    &  \multicolumn{2}{c}{White} \\  \hline
\textbf{}                     & Female\textsubscript{pred}    & Male\textsubscript{pred}      & Female\textsubscript{pred}                   & Male\textsubscript{pred}                 \\
Female\textsubscript{ground}           & 0.98          & 0.02          & \textbf{0.98}                 & 0.02                      \\
Male\textsubscript{ground}             & 0.08          & 0.92          & 0.02                          & \textbf{0.98}            
\end{tabular}}
\label{accuracy}
\end{table}

The overview shows that all metadata categories (gender, race, age) existing in the initial dataset were recognizable by the algorithm. Not only were social constructs such as gender and race inferred by the algorithm, but so did age play a big role in how it saw individuals and how it described them. Furthermore, the algorithm inferred an individual's gender in 15\% of the labels, assigning a non-binary gender to them (e.g. trans) 28 times. It also returned labels about the sexual orientation of individuals (e.g. gay, lesbian) 44 times. This shows that in the algorithm's views and culture people were neither binary gendered nor exclusively heterosexual. Nevertheless, it also means that the algorithm made assumptions about individuals that are not dependent on one's appearance, providing information about how the algorithm evaluated gender and sexuality in its social reality.


\begin{figure}[t]

   \includegraphics[width=1\linewidth,trim={0 0 0.3cm 0},clip]{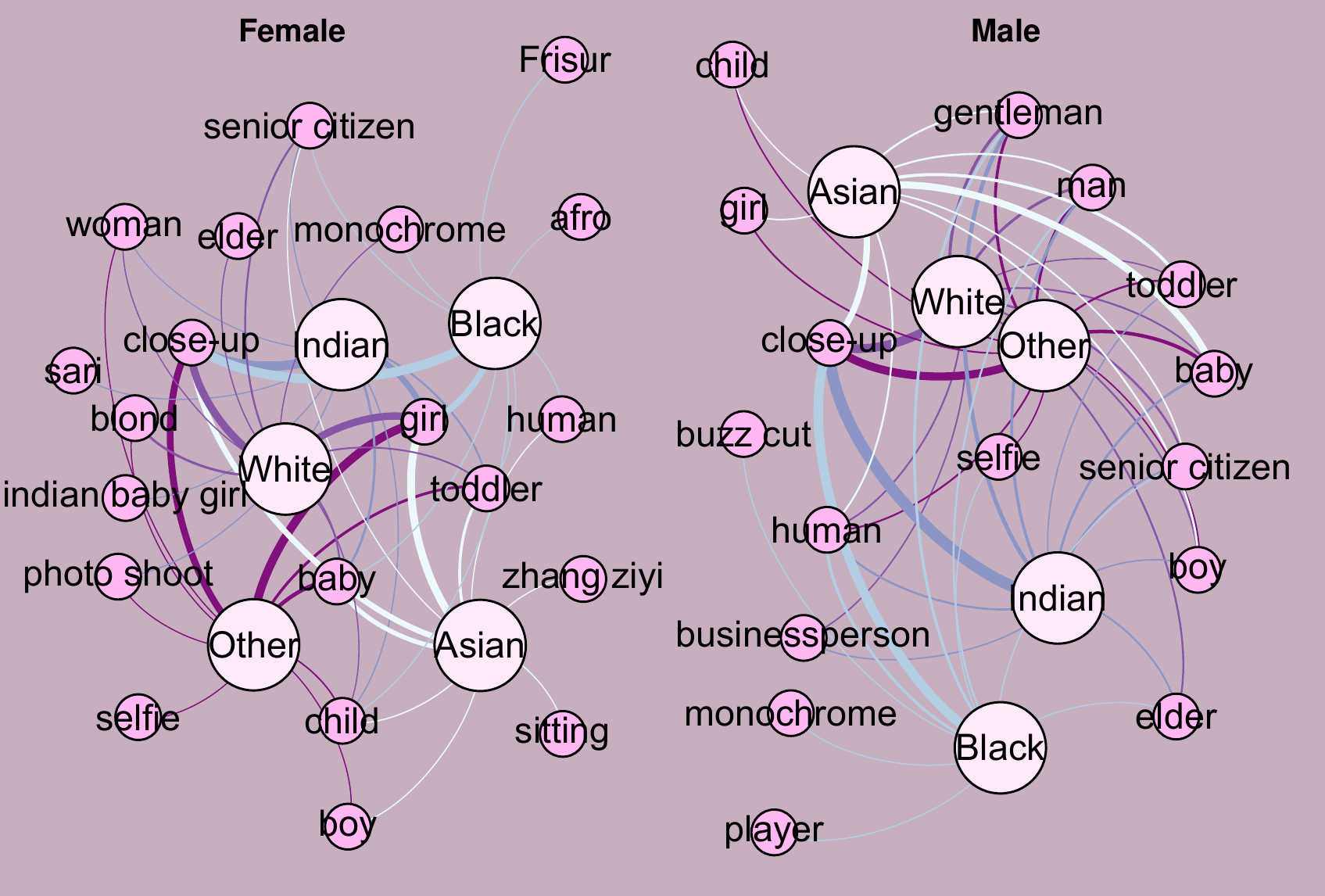}
\caption{The above networks show the algorithm's top generated labels for each gender were distributed across racial groups. The width of each connection correspond to the number of images containing the label for a gender-race group divided by the total number of images for the gender-race group in the dataset.}
   \label{allabels}
\end{figure}
Besides the algorithm's ability to recognize social constructs, it also adapted the language and content of the generated labels according to them. Especially for the language of labels, the assumed racial group of an individual played a significant role. As figure \ref{language} shows, the majority of labels were in English (between 70\% to 90\% depending on the category). Nevertheless, when the picture included someone classified as Asian in the initial dataset, the chance of returning a label in Chinese was significantly higher (10\% of the labels). The same applies for pictures of individuals classified as Indian (6\% of labels in hindu) or individuals from the 'other' category, which included individuals with Latinx origin (6\% of labels in Spanish). This result demonstrates that the algorithm viewed language as an element of the respective culture of individuals.

Although the algorithm adjusted its inferences according to people's gender and race, this was not always performed successfully. Table \ref{accuracy} illustrates the agreement of the initial dataset and algorithm's predictions about people's gender by racial group. As is visible, the algorithm recognized female individuals easier, classifying them as such with a higher accuracy. Furthermore, classification agreement varied between racial groups, with white individuals being classified correctly the most. In contrast, male Asian individuals were misclassified by the highest degree.  Biases are reinforced in the diversity of labels the algorithm assigned to each social group. The analysis showed that the algorithm, in terms of entropy, generated more diverse labels for White men (5.9) followed by White women \& Black men (5.6), followed by Black women (5.5), Indian men, Indian women (5.27), other men (5), Asian men (4.9) and Other women (4.7). These results provide the first evidence that the algorithm treated individuals from different social groups unequally, with the signs of a predilection for men and the White race, possessing a bias that has been found in other algorithmic systems as well \cite{buolamwini2018gender,raji2019actionable}.

The above results already provide significant information about how the algorithm viewed individuals of different genders, ages, and races (RQ1), as well as properties of algorithm's culture (RQ2). The algorithm (re)produced a society categorized by race, gender, and age, in which sexuality is an existent element, and language is seen as a cultural property. Furthermore, the algorithm's culture contained a hierarchical structure among genders and races, as seen by the above biases. In the following, we illustrate further features of this hierarchy.

\subsection{Hierarchical perceptions and attitudes}

\begin{table}[t]
\caption{Logistic regression results. Dependent Variable levels comprised 39 adjectives belonging to six categories: attractiveness, character, feature, sentiment, size, status. Independent variables included race, gender, and age. Model baseline was white-male, and reported results show statistically significant variables (p $\leq$ 0.05) and the sign of the estimator (positive/negative). More detailed results can be found in the appendix. }
\resizebox{3.3in}{!}{\begin{tabular}{lr|l}
\textbf{Category}              & \textbf{Adjective} & \textbf{Significant Effects (p$\leq$0.05)}                                 \\ \hline \hline 

 & Beautiful          & Age (-); Female (+); Black (+); Indian (+);     \\
                                 & Cute               & Age (-); Asian (-); Indian (+); Other (-)       \\
                                 & Handsome           & Female (-); Black (+)                             \\
                                 & Hot                & Age (-); Female (+); Asian(-); Indian (+);      \\
                                 & Pretty             & Age (-); Female (+)                             \\
     Attractiveness                           & Sexy               & Female + Black (+)                              \\
                                 & Ugly               & Black (+); Other (+)                               \\
                                 & Beauty             & Age (-); Black (+): Female (+)                     \\
                                 & Attractive         &                                                 \\
                                 & Prettiest          &                                                 \\
                                 & Sweet              & Age (-)                                         \\
                                 & Smart              & Female (-); Age (-)                               \\ \hline
    & Sweet              & Age (-)                                         \\
                                 & Nice               & Asian (+)                                       \\
                                 & Kind               & Age (-); Black (-); Asian (-); Other (-)        \\
        Character                            & Bad                & Black (+); Indian (+)                           \\
                                 & Good               & Female(-); Asian (-); Indian (-)                      \\
                                 & Cool               & Female(-)                                                \\
                                 & Smart              & Female(-); Age (-)                               \\ \hline
       & Dark               & Black (+)                                       \\
                                 & Light              & Black (+)                                       \\
                                 & Natural            & Female(+); Black (+)                              \\
    Feature                             & Human              & Age (+); Female (-); Black (+); Indian (+); Asian (+) \\
                                 & Average            &  Age (+)                                               \\
                                 & Perfect            &                                                 \\
                                 & Normal             &  Asian (+)                                                \\ \hline
       & Happy              & Age (+)                                         \\
     Sentiment                              & Sad                & Black (+); Asian (+)                               \\
                                 & Angry              &                                                 \\
                                 & Confused           &                                                 \\ \hline
           & Big                & Age (-)                                       \\
                                 & Small              & Age (-)                                       \\
                                 & Thin               & Female (+); Indian (-)                 \\
          Size                        & Fat                & Black (+)                                       \\
                                 & Chubby             & Age (-)                                        \\
                                 & Slim               & Black (+)                                       \\
                                 & Skinny             & Age (+)                                        \\
                                 & Strong             &                                                 \\ \hline
        & Classy             & Age (+)                                         \\
                                 & Traditional        & Female (+)                               \\
             Status                      & Tribal             &                                        \\
                                 & Rich               & Age (+); Asian (-); Indian (-)                 \\
                                                                     & Poor               & Female (+);  Indian (+)                 \\

\end{tabular}}
\label{regression}
\end{table}

The tendency of the algorithm to behave differently based on race and gender becomes clearer when analyzing the distribution of top labels. The networks in figure \ref{allabels} show how the top labels for each gender were distributed across racial groups. The width of each connection shows the strength of labels' association within each group. In the Male network space, we can see the language shared across the races included ``close up'', ``gentleman'' and ``man'', while the Female network space shows that women were typically labeled as ``girl'' across all ages, a troublesome term when used in the wrong social context \cite{fogarty_2019}. Asian men were also often misclassified as ``girls'', a result of the algorithmic bias already discussed above. The networks show that labels related to age were also highly prevalent, with those about children and babies appearing most often for Indian, Asians and Other for both genders. On the other hand, labels associated with the elderly such as "elders" and ``senior citizen'', had higher weights for White men and women. These differences were nevertheless a result of the data distribution in our prompts, and not of the algorithm's tendency to associate specific races with these terms.  Hair was a particularly relevant label for both black men and women, with words such as ``afro'', ``buzz-cut'' and ``frisur'' (German for haircut) being some of their most common attributes, an association replicating the actual socio-political nature of hair for black people \cite{okazawa1987black, dash2006black}. The fact that one of the top labels was in German can be attributed to the IP we used, providing evidence that the algorithm adapted its language based on its knowledge about our location, which is one of the personalization features that the google search engine uses \cite{barysevich_2018}. Hair was also a distinguishing feature for the algorithm when shown white females, often simplifying their appearance under the term ``blond'', another term that carries a socially ambiguous meaning, often seen as positive while other times associated with various negative stereotypes \cite{sherrow2006encyclopedia}. Results also suggest that Black men were frequently labeled as "Player" while Asian and White men as "businessperson", assuming stereotypes in their respective careers \cite{tyler2001biased, paek2003racial}. No such career or action oriented labels appear for women (with the exception of ``sitting'', associated with Asian women, attributed to the initial dataset). Finally, the objectively assumed truth of ``Human'' was more closely related to Black Men and Black \& Asian Women. In general, the word ``human'' was associated with images featuring mostly non-white men. This classification, although technically accurate, suggests that, for the algorithm, ``human'' was implicit in whiteness but not in other races, especially for men. Furthermore, the label ``human'' is an oversimplification, which is a structural element of stereotyping as a human heuristic \cite{mccauley1980stereotyping}.

The multiple logistic regression results (table \ref{regression} and appendix) contribute to the deeper understanding of algorithm's cultural hierarchy that appeared in the findings above in form of biases and stereotypes. The model showed that the algorithm associated character related labels more with men (such as ``smart'' and ``good'') and associations to white individuals were generally more positive. These conceptions agree with the ones existing in a large part of the society and have been produced by historical sociopolitical asymmetries \cite{jaxon2019acquisition, dovidio1986racial}. For example, Black people were associated negatively with kindness and positively with the label "bad", while Asians and Indians were labeled significantly less ``good'' in comparison to White. Similar to the character related labels, in the sentiment category the term "sad" was associated more with Blacks and Asians, while size related labels were more prevalent for younger people, Black individuals and women. Again, the fact that the algorithm associated appearance related adjectives much more to women and non-white individuals is a result of algorithm's hierarchical perceptions, since the focus on individuals' appearance is tightly associated to discrimination \cite{zebrowitz1996physical, johnson2017facial}. Furthermore, the algorithm distributed status related labels in a similar way, associating status label ``rich'' much less with Asians and Indians.

\begin{figure}[t]
  \centering
  \includegraphics[width=0.45\textwidth]{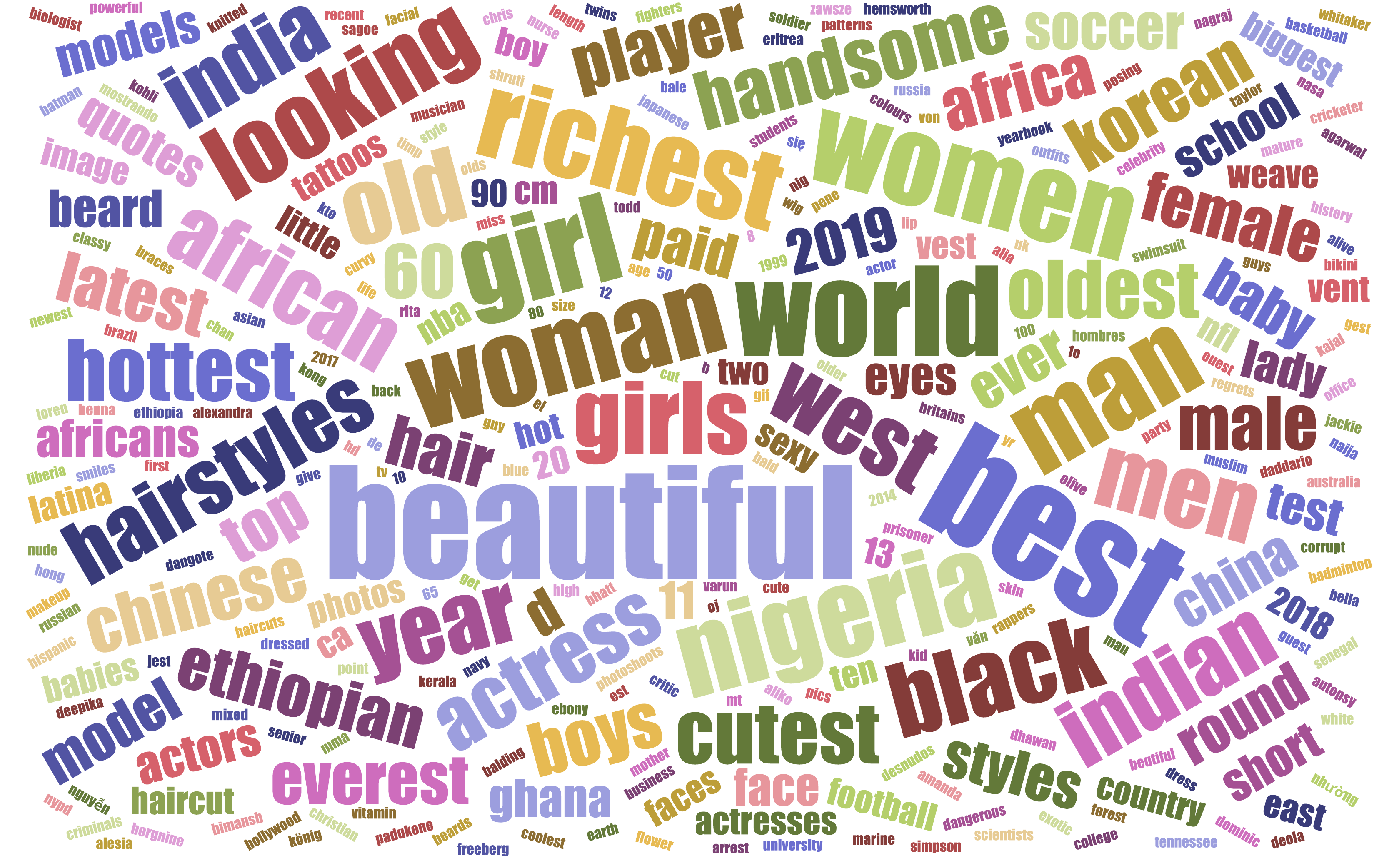}
  \caption{Word cloud of superlative terms used in the labels}
  \label{superlativ}
\end{figure}

\begin{figure}[t]
  \centering
\includegraphics[width=0.4\textwidth]{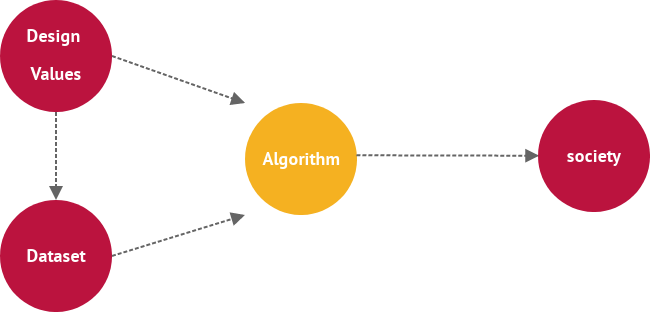}
  \caption{Cultural reproduction within the field of the google search by image algorithm. The algorithm reproduces a culture existing in the dataset it was trained on, and the culture of its owners and designers. Furthermore, the culture of the dataset itself is a result of owners' and designers' decisions. This culture is further replicated in society, by the individuals use of the search engine.}
  \label{field}
\end{figure}

\begin{figure*}[t]
  \centering
\includegraphics[width=1\textwidth]{images/AllTech_compressed.pdf}
  \caption{Labels assigned by the google search by image algorithm to women and men in the top of the techno-hierarchical ladder. The number embedded in colored circles corresponds to the number of times the label in the respective color was associated to an individual. Since the algorithm associated much more labels to men, for simplicity reasons we did not connect all of them to individuals in the visualization.}

  \label{priviledge}
\end{figure*}

A characteristic part of the algorithm's culture was its perception of attractiveness.  The algorithm associated terms from the "attractiveness" category more with women than with men (with the lone exception of the label ``Handsome'') and with younger ages. Furthermore, regardless of whether positive or negative, the algorithm associated labels around attractiveness with non-white races (e.g. ``beautiful'', ``ugly''), once again confirming its focus on the appearance of these races. This persistent association of attractiveness with non-white individuals was not a coincidence. This can not only be seen in terms such as ``hot'' or ``sexy'' in the logistic regression results, but also when looking at the labeled dataset, where images tagged "hot" were mostly non-white women. This finding was also reflected in labels in the dataset that were expressed in superlative forms. As seen in the word cloud in figure \ref{superlativ}, labels such as "best african beautiful girls" and "hottest hong kong women" were really common, whereas labels including the white race were non-existent, suggesting a theme of exoticization towards non-white individuals. This behavior of the algorithm again mirrors perceptions existing in western societies, where female non-whites are often exoticized and sexualized \cite{sue2007racial,forrest2015differences}.

In terms of age, the algorithm associated different groups of adjectives to younger and older people. Positive attractiveness terms such as beautiful, cute, pretty and sweet were associated to younger individuals. The same applied to specific size related terms (big, small, chubby). These algorithmic attitudes conform with general ageist tendencies related to beauty in the society, both on its idealization and oppressive aspects \cite{veresiu2021advanced}. In contrast, concepts revealing well-being such as classy, rich and happy, together with adjectives revealing triviality (average, human) were associated more to older populations.  This shows that the algorithm evaluated age in a generally complex manner. On the one hand, it idealized the appearance of younger individuals, while it ignored or reduced the ones of older people. On the other hand, the algorithm ``recognized'' in older individuals' portraits specific socially enticing values, such as happiness and social status.

The above analysis provided further evidence of the algorithm's culture (RQ2) and of its perceptions on age, gender and race (RQ1). Not only does its culture include a clear hierarchical structure stemming from the white patriarchy, but the algorithm also reproduced behaviors and generated inferences that discriminate and type-cast individuals. Moreover, the algorithm possesses a clear conception of beauty, which strongly aligns with its hierarchical view that white and male is normal.

\subsection{Critical confrontation}

The described white patriarchy, the exoticization of female appearance, the idealization of the young body, the hierarchy of masculinity, the ageist associations and the trivialization of non-white and older populations is not something that the algorithm conceives and produces on its own. The above structures are what Bourdieu would call the \textit{habitus} of a social subject. They are the dispositions, perceptions and attitudes that an individual holds and comprise a \textit{structured and structuring structure} \cite{bourdieu1990other}. They are a structure, because they emerge systematically and regularly through algorithm's inferences. They are structured, because they were learned by the algorithm through a formal training process on a specific dataset, based on designers' and creators' incentives. They are structuring, because they shape the present and future practices of the individuals who use the algorithm. The algorithm's habitus, therefore, is per se relational, and was created and functions within a \textit{field} \cite{maton2014habitus}, a space which includes the individuals and institutions who shaped and are shaped by the algorithm (figure \ref{field}). 

In this field, cultural reproduction is a constitutive part of it. In our case, the algorithm externalized a specific culture through its inferences, which are constantly shown and prescribed to the users who use the search engine. Since the algorithm's habitus is relational, this culture was prescribed to the algorithm by its owners, designers, and training dataset. Similarly, the culture of the training dataset was based on the perceptions, choices and incentives of its designers and owners, who decided to use a set of labels that contained a strong social hierarchy, ignoring its problematic nature. In this way, Bourdieu's theory actually allows us to draw a direct connection between algorithmic inferences, the society and the algorithm's designers and owners. We are able to lay the foundations for impact assessment, since we detect specific social groups that are discriminated by the algorithm, and explain the nature of the discrimination based on social-theoretic knowledge. Furthermore, we are able to formulate statements of accountability, as the framework assigns direct responsibility to the algorithms' owners and designers for any detected effects in the society.

Since Bourdieu's theory draws a direct connection between algorithmic effects and its owners, the ethnographic practice of studying up allows us to challenge the power of powerful individuals in the tech industry to generate culture and to criticize them for their choices. We showed the algorithm the faces of 9 men and women high in the techo-hierarchical ladder ten times, after stripping them of their privilege, and evaluated the answers returned. The findings, found in figure \ref{priviledge}, show that the algorithm placed individuals in the previously described sociocultural reality, creating biased representations of them. First, the algorithm produced much more diverse labels for men than to women (45 vs 13 unique labels). Women were usually described as ``blond'' and ``girl'' (81 \% of the times), while the most frequent label for men was ``gentleman'' (30 \% of the time), with no other label appearing more than 5 times.  As it is directly visible, labels were much more positive for men than for women. Besides the troublesome use of terms such as ``blond'' and ``girl'', the algorithm described female individuals as spokespersons, commented on the hairstyle of Bozoma St. John, and generated a sexual label for Marissa Mayer. Furthermore, the algorithm recognized women or their profession only in 6\% of the time. On the contrary, the algorithm recognized male individuals or their profession 22\% of the time, and associated them with much more action and career oriented terms, such as ``business person'', ``public speaking'', and ``event''. Male labels also had much more positive qualities in them, such as ``fun'' and ``talented.'' The above results demonstrate that the algorithm treated individuals differently based on their gender, confirming that even in the case of powerful actors, the algorithm's biased perceptions persisted. Furthermore, regardless of gender, the algorithm failed to understand individuals behavior accurately. For example, the algorithm returned the label "singing", because individuals were holding a microphone. Moreover, the algorithm characterized Jeff Bezos as a nerd, and saw Eric Schmidt as a police officer. 

By studying up powerful individuals, and making them the object of algorithmic investigation (RQ3), we managed to point out the algorithm's culture, predispositions, and vulnerabilities in a provocative way. Not only did the algorithm externalize its biased views about the world on individuals, discriminating women, but also exhibited frequent misclassifications. On top of that, by changing the population under investigation and focusing upward, we were able to show the extreme lack of racial and gender diversity in high positions in the tech industry. Most individuals are white, and women had professions of lower importance in the hierarchy. Overall, by performing our analysis, we uncovered an important part of the algorithm's culture, as well as of the total system that the algorithm is an element of.

\section{Scientific and Design Impetus}

\begin{figure*}[t]
\centering
\includegraphics[width=0.95\textwidth]{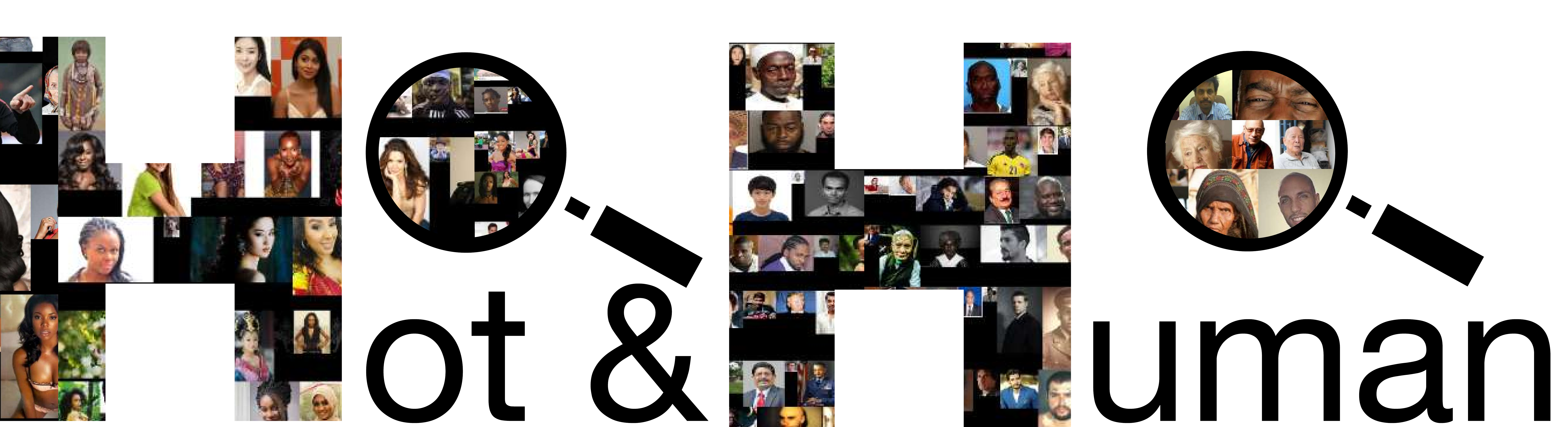}
  \caption{Image cloud for a random sample of images labeled as hot (left) and human (right). In the vast amount of cases the
algorithm labeled as such non-white females and men respectively, exoticizing and oversimplifying individuals.}
  \label{HOTHUMAN}
\end{figure*}

The socio-computational interrogation of the google search by image algorithm uncovered part of the social and cultural structures (re)produced by a main component of the most popular search engine in the world. The analysis revealed that a major commercial algorithmic implementation results in unjust associations, inferences, and predictions, favoring white-male individuals and discriminating against people of other genders and races. Besides describing this phenomenon in-detail, the study provoked key-stakeholders in the tech-industry to reflect on the issue, by making them the object of algorithm's hierarchical conceptions.

The integration of the sociological framework of Bourdieu and the anthropological tactics of Nader allowed us to perform an algorithmic bias study that goes beyond simply auditing the algorithm. The theory of cultural reproduction allowed us to frame algorithmic behavior in terms of the cultural values it mediates and to locate discriminatory practices stemming from the white patriarchy, instead of just reporting mathematical differences between social groups. This knowledge can be useful when performing further studies or auditing procedures that quantify the exact impact of biases that algorithmic systems have on user groups (see e.g. \cite{metaxa2021image, mitchell2019model}). Furthermore, the theory assigned clear-cut responsibility to the algorithm's designers and owners for its problematic behavior by positioning the algorithmic bias study within the social context the algorithm functions, hence promoting visibility and accountability. Similarly, the practice of studying up allowed us to highlight existing power-dynamics in the environments that create these algorithms. It also allowed us to critically confront these environments, making them the object of the algorithm's inferences, underpinning the normative dimension of our study in a more interesting way. We do not claim that we exhaustively applied the above theories in interpreting the google search by image algorithm. Nonetheless, we argue that we showed the value of using the specific or similar social-theoretic frameworks when performing algorithmic bias or auditing studies.

While performing the above contributions, we encountered multiple issues related to the scientific rigour of our analysis, as well as the design vulnerabilities of the investigated algorithmic system.

From a scientific perspective, we faced technical difficulties and ethical dilemmas during study design and evaluation. First, in order to analyze the algorithm's conceptions about race and gender, we had to use a dataset containing images with metadata algorithmically generated, which only included four main races and two genders. This not only automatically erased other races, making our own analysis less diverse, but also the initial data does not correspond to an actual ground truth, but an approximate one. Given this approximation, it was troublesome for us to define the boarders of social constructs of race and gender and to perform the analysis in a robust way. Taking this into consideration, we urge the scientific community to consider developing datasets that present identities as constructed and fluid features, and not as rigid categorizations. In cases that this is not possible, then it is scientifically and ethically more righteous to describe in detail the ethical and practical limitations and consequences of the study. This path is necessary, especially in cases when the actual individual did not provide information on or about themselves. We faced similar issues when using pre-trained language models for lingual entity recognition. Not only did the models not cover all the languages appearing in our dataset, but algorithmic predictions were not perfect either. Scientists should therefore be conscious that existing technical solutions pose serious limitations towards diverse and inclusive research. In addition, diversifying scientific analysis presumes much additional work, which scientists should be willing to perform, by considering at the start of the study who should be included and why. 

From an algorithmic design perspective, the fact that the google search by image algorithm (re)produces the structures of white patriarchy is highly problematic. The oversimplified representation of females and non-white individuals, and the stereotyping and discrimination against them illustrates once again how much should be done by private and public actors toward just algorithmic implementations. Especially when going through the generated labels, we found the
recurring regularity of non-white individuals being presented under the theme \textit{hot \& and human} (figure \ref{HOTHUMAN}). Non-white women, especially the young ones, were exoticized and sexualized, while non-white men were reduced to the self-evident description of being human; the
least information an algorithm could provide about an individual. As described above, the algorithm's behavior mirrors existing social conditions, which are the result of socio-historical and political processes that have generated structural asymmetries. Nevertheless, biases and representations such as the above should not be part of one of the most popular algorithmic implementations in the world. Even if the algorithm learned and returned queries corresponding to associations that users previously made, designers should generate frameworks that prevent the harm of individuals regardless of their social group; not doing may suggest that the above described issues are not of central concern to them. The mitigation of harms could be achieved by making equal predictions for individuals regardless their gender, race, or age, by avoiding oversimplifications and historical stereotypes associated with groups, and providing ethical guidelines that clearly state the result categories that search algorithms should yield. A prerequisite for this is the understanding of the social context in which an algorithm is employed, how it will be used, and how the algorithm will impact the diverse user population \cite{selbst2019fairness}. There is still much to be done towards just algorithmic implementations, and public and private actors should work toward this direction using transparent and accountable means.

\section{Conclusion}

In this study, we performed a socio-computational interrogation of the google search by image algorithm, a main component of the google search engine. Drawing from prior theoretical work, we created a framework for understanding and evaluating the algorithm's culture. By using computational and qualitative tools, we found that the algorithm reproduced structures stemming from white patriarchal society, discriminating and stereotyping females and non-white social groups. As a scientific and design provocation, we studied up key stakeholders in the tech industry, stripping them of their privileges and showing how the algorithm places them within its socioalgorithmic reality, generating biased representations of them. Based on the results, we discussed scientific and design implications and solutions towards just and inclusive socioalgorithmic systems.


\bibliography{aaa21}

\begin{thebibliography}{70}
\providecommand{\natexlab}[1]{#1}
\providecommand{\url}[1]{\texttt{#1}}
\providecommand{\urlprefix}{URL }
\expandafter\ifx\csname urlstyle\endcsname\relax
  \providecommand{\doi}[1]{doi:\discretionary{}{}{}#1}\else
  \providecommand{\doi}{doi:\discretionary{}{}{}\begingroup
  \urlstyle{rm}\Url}\fi

\bibitem[{Abebe et~al.(2020)Abebe, Barocas, Kleinberg, Levy, Raghavan, and
  Robinson}]{abebe2020roles}
Abebe, R.; Barocas, S.; Kleinberg, J.; Levy, K.; Raghavan, M.; and Robinson,
  D.~G. 2020.
\newblock Roles for computing in social change.
\newblock In \emph{Proceedings of the 2020 Conference on Fairness,
  Accountability, and Transparency}, 252--260.

\bibitem[{Abramson et~al.(2018)Abramson, Joslyn, Rendle, Garrett, and
  Dohan}]{abramson2018promises}
Abramson, C.~M.; Joslyn, J.; Rendle, K.~A.; Garrett, S.~B.; and Dohan, D. 2018.
\newblock The promises of computational ethnography: improving transparency,
  replicability, and validity for realist approaches to ethnographic analysis.
\newblock \emph{Ethnography} 19(2): 254--284.

\bibitem[{Alvarado and Humphreys(2017)}]{alvarado2017big}
Alvarado, R.; and Humphreys, P. 2017.
\newblock Big data, thick mediation, and representational opacity.
\newblock \emph{New Literary History} 48(4): 729--749.

\bibitem[{Barabas et~al.(2020)Barabas, Doyle, Rubinovitz, and
  Dinakar}]{barabas2020studying}
Barabas, C.; Doyle, C.; Rubinovitz, J.; and Dinakar, K. 2020.
\newblock Studying up: reorienting the study of algorithmic fairness around
  issues of power.
\newblock In \emph{Proceedings of the 2020 Conference on Fairness,
  Accountability, and Transparency}, 167--176.

\bibitem[{Barocas, Hardt, and Narayanan(2017)}]{barocas_hardt_narayanan}
Barocas, S.; Hardt, M.; and Narayanan, A. 2017.
\newblock \urlprefix\url{https://fairmlbook.org/}.

\bibitem[{Barrett and Kreiss(2019)}]{barrett2019platform}
Barrett, B.; and Kreiss, D. 2019.
\newblock Platform transience: changes in Facebook's policies, procedures, and
  affordances in global electoral politics.
\newblock \emph{Internet Policy Review} 8(4): 1--22.

\bibitem[{Barysevich(2018)}]{barysevich_2018}
Barysevich, A. 2018.
\newblock 5 Ways to Optimize for Personalized Search.
\newblock
  \urlprefix\url{https://www.searchenginejournal.com/5-ways-to-optimize-for-personalized-search/263499/}.

\bibitem[{Blodgett et~al.(2020)Blodgett, Barocas, Daum{\'e}~III, and
  Wallach}]{blodgett2020language}
Blodgett, S.~L.; Barocas, S.; Daum{\'e}~III, H.; and Wallach, H. 2020.
\newblock Language (Technology) is Power: A Critical Survey of" Bias" in NLP.
\newblock \emph{arXiv preprint arXiv:2005.14050} .

\bibitem[{Bourdieu(1987)}]{bourdieu1987makes}
Bourdieu, P. 1987.
\newblock What makes a social class? On the theoretical and practical existence
  of groups.
\newblock \emph{Berkeley journal of sociology} 32: 1--17.

\bibitem[{Bourdieu(1990)}]{bourdieu1990other}
Bourdieu, P. 1990.
\newblock \emph{In other words: Essays towards a reflexive sociology}.
\newblock Stanford University Press.

\bibitem[{Bourdieu and Passeron(1990)}]{bourdieu1990reproduction}
Bourdieu, P.; and Passeron, J.-C. 1990.
\newblock \emph{Reproduction in education, society and culture}, volume~4.
\newblock Sage.

\bibitem[{Buolamwini and Gebru(2018)}]{buolamwini2018gender}
Buolamwini, J.; and Gebru, T. 2018.
\newblock Gender shades: Intersectional accuracy disparities in commercial
  gender classification.
\newblock In \emph{Conference on fairness, accountability and transparency},
  77--91.

\bibitem[{Burrell(2016)}]{burrell2016machine}
Burrell, J. 2016.
\newblock How the machine ‘thinks’: Understanding opacity in machine
  learning algorithms.
\newblock \emph{Big Data \& Society} 3(1): 2053951715622512.

\bibitem[{Cave and {\'O}h{\'E}igeartaigh(2018)}]{cave2018ai}
Cave, S.; and {\'O}h{\'E}igeartaigh, S.~S. 2018.
\newblock An AI race for strategic advantage: rhetoric and risks.
\newblock In \emph{Proceedings of the 2018 AAAI/ACM Conference on AI, Ethics,
  and Society}, 36--40.

\bibitem[{Christin(2020)}]{christin2020ethnographer}
Christin, A. 2020.
\newblock The ethnographer and the algorithm: beyond the black box.
\newblock \emph{Theory and Society} 1--22.

\bibitem[{Cloud(2020)}]{google}
Cloud, G. 2020.
\newblock Cloud Translation API.
\newblock
  \urlprefix\url{https://cloud.google.com/translate/docs/reference/rest}.

\bibitem[{Dash(2006)}]{dash2006black}
Dash, P. 2006.
\newblock Black hair culture, politics and change.
\newblock \emph{International journal of inclusive education} 10(1): 27--37.

\bibitem[{Denton et~al.(2020)Denton, Hanna, Amironesei, Smart, Nicole, and
  Scheuerman}]{denton2020bringing}
Denton, E.; Hanna, A.; Amironesei, R.; Smart, A.; Nicole, H.; and Scheuerman,
  M.~K. 2020.
\newblock Bringing the People Back In: Contesting Benchmark Machine Learning
  Datasets.
\newblock \emph{arXiv preprint arXiv:2007.07399} .

\bibitem[{Diaz(2008)}]{Diaz2008}
Diaz, A. 2008.
\newblock \emph{Through the Google Goggles: Sociopolitical Bias in Search
  Engine Design}, 11--34.
\newblock Berlin, Heidelberg: Springer Berlin Heidelberg.
\newblock ISBN 978-3-540-75829-7.
\newblock \doi{10.1007/978-3-540-75829-7_2}.
\newblock \urlprefix\url{https://doi.org/10.1007/978-3-540-75829-7_2}.

\bibitem[{D'Ignazio and Klein(2020)}]{d2020data}
D'Ignazio, C.; and Klein, L.~F. 2020.
\newblock \emph{Data feminism}.
\newblock MIT Press.

\bibitem[{Dixon-Rom{\'a}n(2016)}]{dixon2016diffractive}
Dixon-Rom{\'a}n, E.~J. 2016.
\newblock Diffractive possibilities: Cultural studies and quantification.
\newblock \emph{Transforming Anthropology} 24(2): 157--167.

\bibitem[{Dovidio, Evans, and Tyler(1986)}]{dovidio1986racial}
Dovidio, J.~F.; Evans, N.; and Tyler, R.~B. 1986.
\newblock Racial stereotypes: The contents of their cognitive representations.
\newblock \emph{Journal of Experimental Social Psychology} 22(1): 22--37.

\bibitem[{Elish and Boyd(2018)}]{elish2018situating}
Elish, M.~C.; and Boyd, D. 2018.
\newblock Situating methods in the magic of Big Data and AI.
\newblock \emph{Communication monographs} 85(1): 57--80.

\bibitem[{Fogarty(2019)}]{fogarty_2019}
Fogarty, M. 2019.
\newblock How to call someone 'girl' without seeming sexist.
\newblock
  \urlprefix\url{https://www.washingtonpost.com/posteverything/wp/2014/07/16/how-to-call-someone-girl-without-seeming-like-a-raging-sexist/}.

\bibitem[{Forrest-Bank and Jenson(2015)}]{forrest2015differences}
Forrest-Bank, S.; and Jenson, J.~M. 2015.
\newblock Differences in experiences of racial and ethnic microaggression among
  Asian, Latino/Hispanic, Black, and White young adults.
\newblock \emph{J. Soc. \& Soc. Welfare} 42: 141.

\bibitem[{Forsythe(1999)}]{forsythe1999ethics}
Forsythe, D.~E. 1999.
\newblock Ethics and politics of studying up in technoscience.
\newblock \emph{Anthropology of Work Review} 20(1): 6--11.

\bibitem[{Gillespie(2014)}]{gillespie2014relevance}
Gillespie, T. 2014.
\newblock The relevance of algorithms.
\newblock \emph{Media technologies: Essays on communication, materiality, and
  society} 167(2014): 167.

\bibitem[{Haines(2017)}]{haines2017towards}
Haines, J.~K. 2017.
\newblock Towards Multi-Dimensional Ethnography.
\newblock In \emph{Ethnographic Praxis in Industry Conference Proceedings}, 1,
  127--141. Wiley Online Library.

\bibitem[{Haraway(1988)}]{haraway1988situated}
Haraway, D. 1988.
\newblock Situated knowledges: The science question in feminism and the
  privilege of partial perspective.
\newblock \emph{Feminist studies} 14(3): 575--599.

\bibitem[{Honnibal and Montani(2017)}]{spacy2}
Honnibal, M.; and Montani, I. 2017.
\newblock {spaCy 2}: Natural language understanding with {B}loom embeddings,
  convolutional neural networks and incremental parsing.
\newblock To appear.

\bibitem[{Hsu(2014)}]{hsu2014digital}
Hsu, W.~F. 2014.
\newblock Digital ethnography toward augmented empiricism: A new methodological
  framework.
\newblock \emph{Journal of Digital Humanities} 3(1): 3--1.

\bibitem[{Jaxon et~al.(2019)Jaxon, Lei, Shachnai, Chestnut, and
  Cimpian}]{jaxon2019acquisition}
Jaxon, J.; Lei, R.~F.; Shachnai, R.; Chestnut, E.~K.; and Cimpian, A. 2019.
\newblock The acquisition of gender stereotypes about intellectual ability:
  Intersections with race.
\newblock \emph{Journal of Social Issues} 75(4): 1192--1215.

\bibitem[{Johnson and King(2017)}]{johnson2017facial}
Johnson, B.~D.; and King, R.~D. 2017.
\newblock Facial profiling: Race, physical appearance, and punishment.
\newblock \emph{Criminology} 55(3): 520--547.

\bibitem[{K{\"a}rkk{\"a}inen and Joo(2019)}]{karkkainen2019fairface}
K{\"a}rkk{\"a}inen, K.; and Joo, J. 2019.
\newblock Fairface: Face attribute dataset for balanced race, gender, and age.
\newblock \emph{arXiv preprint arXiv:1908.04913} .

\bibitem[{Markham(2016)}]{markham2016ethnography}
Markham, A.~N. 2016.
\newblock Ethnography in the digital internet era.
\newblock \emph{Denzin NK \& Lincoln YS, Sage handbook of qualitative research,
  Thousands Oaks, CA: Sage Publications} 650--668.

\bibitem[{Maton(2014)}]{maton2014habitus}
Maton, K. 2014.
\newblock Habitus.
\newblock In \emph{Pierre Bourdieu}, 60--76. Routledge.

\bibitem[{McCauley, Stitt, and Segal(1980)}]{mccauley1980stereotyping}
McCauley, C.; Stitt, C.~L.; and Segal, M. 1980.
\newblock Stereotyping: From prejudice to prediction.
\newblock \emph{Psychological Bulletin} 87(1): 195.

\bibitem[{McLuhan and Fiore(1967)}]{mcluhan1967medium}
McLuhan, M.; and Fiore, Q. 1967.
\newblock The medium is the message.
\newblock \emph{New York} 123: 126--128.

\bibitem[{Metaxa et~al.(2021)Metaxa, Gan, Goh, Hancock, and
  Landay}]{metaxa2021image}
Metaxa, D.; Gan, M.~A.; Goh, S.; Hancock, J.; and Landay, J.~A. 2021.
\newblock An Image of Society: Gender and Racial Representation and Impact in
  Image Search Results for Occupations.
\newblock \emph{Proceedings of the ACM on Human-Computer Interaction} 5(CSCW1):
  1--23.

\bibitem[{Mitchell et~al.(2019)Mitchell, Wu, Zaldivar, Barnes, Vasserman,
  Hutchinson, Spitzer, Raji, and Gebru}]{mitchell2019model}
Mitchell, M.; Wu, S.; Zaldivar, A.; Barnes, P.; Vasserman, L.; Hutchinson, B.;
  Spitzer, E.; Raji, I.~D.; and Gebru, T. 2019.
\newblock Model cards for model reporting.
\newblock In \emph{Proceedings of the conference on fairness, accountability,
  and transparency}, 220--229.

\bibitem[{Mittelstadt et~al.(2016)Mittelstadt, Allo, Taddeo, Wachter, and
  Floridi}]{mittelstadt2016ethics}
Mittelstadt, B.~D.; Allo, P.; Taddeo, M.; Wachter, S.; and Floridi, L. 2016.
\newblock The ethics of algorithms: Mapping the debate.
\newblock \emph{Big Data \& Society} 3(2): 2053951716679679.

\bibitem[{Murray-Rust et~al.(2019)Murray-Rust, Gorkovenko, Burnett, and
  Richards}]{murray2019entangled}
Murray-Rust, D.; Gorkovenko, K.; Burnett, D.; and Richards, D. 2019.
\newblock Entangled Ethnography: Towards a collective future understanding.
\newblock In \emph{Proceedings of the Halfway to the Future Symposium 2019},
  1--10.

\bibitem[{Nader(1972)}]{nader1972up}
Nader, L. 1972.
\newblock Up the anthropologist: perspectives gained from studying up.
\newblock \emph{ERIC} .

\bibitem[{Neyland(2019)}]{neyland2019everyday}
Neyland, D. 2019.
\newblock \emph{The everyday life of an algorithm}.
\newblock Springer Nature.

\bibitem[{Noble(2018)}]{noble2018algorithms}
Noble, S.~U. 2018.
\newblock \emph{Algorithms of oppression: How search engines reinforce racism}.
\newblock nyu Press.

\bibitem[{Obermeyer and Mullainathan(2019)}]{obermeyer2019dissecting}
Obermeyer, Z.; and Mullainathan, S. 2019.
\newblock Dissecting racial bias in an algorithm that guides health decisions
  for 70 million people.
\newblock In \emph{Proceedings of the Conference on Fairness, Accountability,
  and Transparency}, 89--89.

\bibitem[{Okazawa-Rey, Robinson, and Ward(1987)}]{okazawa1987black}
Okazawa-Rey, M.; Robinson, T.; and Ward, J.~V. 1987.
\newblock Black women and the politics of skin color and hair.
\newblock \emph{Women \& Therapy} 6(1-2): 89--102.

\bibitem[{Paek and Shah(2003)}]{paek2003racial}
Paek, H.~J.; and Shah, H. 2003.
\newblock Racial ideology, model minorities, and the" not-so-silent partner:"
  Stereotyping of Asian Americans in US magazine advertising.
\newblock \emph{Howard Journal of Communication} 14(4): 225--243.

\bibitem[{Papakyriakopoulos et~al.(2020)Papakyriakopoulos, Hegelich, Serrano,
  and Marco}]{papakyriakopoulos2020bias}
Papakyriakopoulos, O.; Hegelich, S.; Serrano, J. C.~M.; and Marco, F. 2020.
\newblock Bias in word embeddings.
\newblock In \emph{Proceedings of the 2020 Conference on Fairness,
  Accountability, and Transparency}, 446--457.

\bibitem[{Pasquale(2015)}]{pasquale2015black}
Pasquale, F. 2015.
\newblock \emph{The black box society}.
\newblock Harvard University Press.

\bibitem[{Peters and Wendland(2016)}]{peters2016up}
Peters, R.~W.; and Wendland, C. 2016.
\newblock Up the Africanist: the possibilities and problems of ‘studying
  up’in Africa.
\newblock \emph{Critical African Studies} 8(3): 239--254.

\bibitem[{Raji and Buolamwini(2019)}]{raji2019actionable}
Raji, I.~D.; and Buolamwini, J. 2019.
\newblock Actionable auditing: Investigating the impact of publicly naming
  biased performance results of commercial ai products.
\newblock In \emph{Proceedings of the 2019 AAAI/ACM Conference on AI, Ethics,
  and Society}, 429--435.

\bibitem[{Rothe, Timofte, and Van~Gool(2015)}]{rothe2015dex}
Rothe, R.; Timofte, R.; and Van~Gool, L. 2015.
\newblock Dex: Deep expectation of apparent age from a single image.
\newblock In \emph{Proceedings of the IEEE international conference on computer
  vision workshops}, 10--15.

\bibitem[{Sandvig et~al.(2014)Sandvig, Hamilton, Karahalios, and
  Langbort}]{sandvig2014auditing}
Sandvig, C.; Hamilton, K.; Karahalios, K.; and Langbort, C. 2014.
\newblock Auditing algorithms: Research methods for detecting discrimination on
  internet platforms.
\newblock \emph{Data and discrimination: converting critical concerns into
  productive inquiry} 22: 4349--4357.

\bibitem[{Seaver(2017)}]{seaver2017algorithms}
Seaver, N. 2017.
\newblock Algorithms as culture: Some tactics for the ethnography of
  algorithmic systems.
\newblock \emph{Big Data \& Society} 4(2): 2053951717738104.

\bibitem[{Seaver(2018)}]{seaver2018should}
Seaver, N. 2018.
\newblock What should an anthropology of algorithms do?
\newblock \emph{Cultural anthropology} 33(3): 375--385.

\bibitem[{Selbst et~al.(2019)Selbst, Boyd, Friedler, Venkatasubramanian, and
  Vertesi}]{selbst2019fairness}
Selbst, A.~D.; Boyd, D.; Friedler, S.~A.; Venkatasubramanian, S.; and Vertesi,
  J. 2019.
\newblock Fairness and abstraction in sociotechnical systems.
\newblock In \emph{Proceedings of the Conference on Fairness, Accountability,
  and Transparency}, 59--68.

\bibitem[{Sherrow(2006)}]{sherrow2006encyclopedia}
Sherrow, V. 2006.
\newblock \emph{Encyclopedia of hair: A cultural history}.
\newblock Greenwood Publishing Group.

\bibitem[{Stich and Colyar(2015)}]{stich2015thinking}
Stich, A.~E.; and Colyar, J.~E. 2015.
\newblock Thinking relationally about studying ‘up’.
\newblock \emph{British Journal of Sociology of Education} 36(5): 729--746.

\bibitem[{Sue et~al.(2007)Sue, Bucceri, Lin, Nadal, and Torino}]{sue2007racial}
Sue, D.~W.; Bucceri, J.; Lin, A.~I.; Nadal, K.~L.; and Torino, G.~C. 2007.
\newblock Racial microaggressions and the Asian American experience.
\newblock \emph{Cultural diversity and ethnic minority psychology} 13(1): 72.

\bibitem[{Tyler~Eastman(2001)}]{tyler2001biased}
Tyler~Eastman, Andrew C.~Billings, S. 2001.
\newblock Biased voices of sports: Racial and gender stereotyping in college
  basketball announcing.
\newblock \emph{Howard Journal of Communications} 12(4): 183--201.

\bibitem[{Urman, Makhortykh, and Ulloa(2021)}]{urman2021matter}
Urman, A.; Makhortykh, M.; and Ulloa, R. 2021.
\newblock The Matter of Chance: Auditing Web Search Results Related to the 2020
  US Presidential Primary Elections Across Six Search Engines.
\newblock \emph{Social science computer review} 08944393211006863.

\bibitem[{van~den Broek(2019)}]{van2019hiring}
van~den Broek, E. 2019.
\newblock Hiring Algorithms: An Ethnography of Fairness in Practice.
\newblock \emph{ICIS 2019 Proceedings} .

\bibitem[{Veresiu and Parmentier(2021)}]{veresiu2021advanced}
Veresiu, E.; and Parmentier, M.-A. 2021.
\newblock Advanced Style Influencers: Confronting Gendered Ageism in Fashion
  and Beauty Markets.
\newblock \emph{Journal of the Association for Consumer Research} 6(2):
  000--000.

\bibitem[{Vincent(2018)}]{vincent_2018}
Vincent, J. 2018.
\newblock Google 'fixed' its racist algorithm by removing gorillas from its
  image-labeling tech.
\newblock
  \urlprefix\url{https://www.theverge.com/2018/1/12/16882408/google-racist-gorillas-photo-recognition-algorithm-ai}.

\bibitem[{Winner(1980)}]{winner1980artifacts}
Winner, L. 1980.
\newblock Do artifacts have politics?
\newblock \emph{Daedalus} 121--136.

\bibitem[{Woodruff et~al.(2018)Woodruff, Fox, Rousso-Schindler, and
  Warshaw}]{woodruff2018qualitative}
Woodruff, A.; Fox, S.~E.; Rousso-Schindler, S.; and Warshaw, J. 2018.
\newblock A qualitative exploration of perceptions of algorithmic fairness.
\newblock In \emph{Proceedings of the 2018 chi conference on human factors in
  computing systems}, 1--14.

\bibitem[{Yang et~al.(2020)Yang, Qinami, Fei-Fei, Deng, and
  Russakovsky}]{yang2020towards}
Yang, K.; Qinami, K.; Fei-Fei, L.; Deng, J.; and Russakovsky, O. 2020.
\newblock Towards fairer datasets: Filtering and balancing the distribution of
  the people subtree in the imagenet hierarchy.
\newblock In \emph{Proceedings of the 2020 Conference on Fairness,
  Accountability, and Transparency}, 547--558.

\bibitem[{Zebrowitz(1996)}]{zebrowitz1996physical}
Zebrowitz, L.~A. 1996.
\newblock Physical appearance as a basis of stereotyping.
\newblock \emph{Stereotypes and stereotyping} 79--120.

\bibitem[{Zhang and Qi(2017)}]{zhifei2017cvpr}
Zhang, Z. S.~Y.; and Qi, H. 2017.
\newblock Age Progression/Regression by Conditional Adversarial Autoencoder.
\newblock In \emph{IEEE Conference on Computer Vision and Pattern Recognition
  (CVPR)}. IEEE.

\end{thebibliography}
\onecolumn

\section{Appendix}

\begin{table}[h]
\fontsize{8}{11}\selectfont
\caption{Multinomial logistic regression results. Significance codes: ** p$\leq$ 0.01, * p $\leq$0.05.}
\begin{tabular}{ccccccccccccc}
\multicolumn{13}{c}{Attractiveness}                                                                                           \\ \hline
         & beautiful & cute  & handsome & hot   & pretty & sexy  & ugly  & beauty & attractive & prettiest & sweet & smart \\ \hline
age       & \textbf{-0.01**}     & \textbf{-0.13**} & -0.01    & \textbf{-0.01**} & \textbf{-0.05**}  & -0.01 & 0.01  & \textbf{-0.04**}  & -0.03      & -0.06     & -0.08 & \textbf{-0.05**} \\
female    & \textbf{2.72**}     & 0.19  & \textbf{-3.85**}    & \textbf{0.40**}  & \textbf{2.14**}   & \textbf{0.99**}  & 0.22  & \textbf{2.53** }  & -0.12      & 3.25      & 1.08  & \textbf{-1.67*} \\
black     & \textbf{1.69**}     & 0.15  & \textbf{0.90**}     & 0.15  & -0.30  & \textbf{0.99**}  & \textbf{2.18**}  & \textbf{1.29**}   & 0.81       & -2.14     & -0.80 & -4.17  \\
asian     & 0.03      & \textbf{-0.83**} & 0.26     & \textbf{-0.78**} & -0.96  & 0.58  & 1.41  & -0.99  & -1.28      & 1.02      & -4.80 & -4.43  \\
indian    & \textbf{0.89**}      & \textbf{0.58**}  & -0.16    & \textbf{0.78**}  & -1.05  & -0.07 & -3.21 & 0.23   & -4.24      & -2.86     & 0.24  & 0.71  \\
other     & 0.33      & \textbf{-0.96**} & 0.15     & 0.08  & 0.81   & -0.24 & \textbf{2.56**}  & -4.20  & -3.55      & -2.42     & -4.04 & -3.54  \\
$\beta_0$     & \textbf{-7.51**}     & \textbf{-2.85**} & \textbf{-6.18**}    & \textbf{-4.47**} & \textbf{-7.49**}  & \textbf{-7.20**} & \textbf{-9.85**} & \textbf{-8.32**}  & \textbf{-6.94**}      & \textbf{-11.21**}    & \textbf{-6.65**} & \textbf{-5.99**}            \\  \hline
\end{tabular}

\hfill \break \newline
\centering
\begin{tabular}{cccccccc}
\multicolumn{8}{c}{Character}                                           \\ \hline
        & sweet & nice   & kind  & bad   & good  & cool  & smart \\ \hline
age    & \textbf{-0.08**} & 0.00   & \textbf{-0.12**} & -0.01 & -0.01 & 0.00  & \textbf{-0.05**} \\
female & 1.08  & 1.42   & 0.16  & -0.32 & \textbf{-0.72**} & \textbf{-2.14*} & \textbf{-1.67*} \\
black  & -0.80 & 1.42   & -1.12 & \textbf{0.98*}  & -0.31 & 0.08  & -4.17 \\
asian  & -4.80 & \textbf{2.02*}   & \textbf{-1.31**} & -4.29 & \textbf{-1.41*} & -3.84 & -4.43 \\
indian & 0.24  & 0.48   & -6.39 & \textbf{1.14*}  & -1.03 & -4.12 & 0.71  \\
other  & -4.04 & -2.60  & \textbf{-2.62**} & 0.73  & -1.09 & -3.01 & -3.54 \\
$\beta_0$  & \textbf{-6.65**} & \textbf{-10.13**} & \textbf{-3.95**} & \textbf{-7.21**} & \textbf{-5.82**} & \textbf{-7.46**} & \textbf{-5.99**}             \\ \hline
\end{tabular}
\begin{tabular}{ccccc}
\multicolumn{4}{c}{Sentiment}             \\ \hline
happy & sad   & angry & confused \\ \hline
\textbf{0.02**}  & -0.01 & -0.02 & -0.06    \\
-0.39 & 0.26  & -0.31 & -0.49    \\
-0.64 & \textbf{1.35**}  & -1.00 & 0.05     \\
0.38  & 0.51  & 0.65  & -0.66    \\
0.16  & \textbf{1.42**}  & 0.87  & -3.38    \\
0.84  & 0.07  & -0.30 & -2.88    \\
\textbf{-7.17**} & \textbf{-7.37**} & \textbf{-7.27**} & \textbf{-6.96**}       \\ \hline
\end{tabular}

\hfill \break \newline
\centering
\begin{tabular}{cccccc}
\multicolumn{6}{c}{Status}           \\  \hline
     & classy & traditional & tribal & rich  & poor\\ \hline
age    & \textbf{0.05*}   & -0.01       & -0.03  & \textbf{0.03**} & 0.00\\
female & 1.07   & \textbf{2.61**}        & 0.82   & -0.41 & \textbf{0.58*}\\
black  & 2.56   & 5.71        & 6.22   & 0.40 & -0.54\\
asian  & -3.05  & 4.18        & -3.00  & \textbf{-0.92*} & -4.22\\
indian & -2.49  & 5.54        & -2.25  & \textbf{-1.36**} & \textbf{2.69**}\\
other  & -1.74  & -2.43       & -1.78  & -0.52 & -3.48\\
$\beta_0$  & \textbf{-13.69**} & \textbf{-13.84**}      & \textbf{-12.41**} & \textbf{-6.98**}  & \textbf{-7.93**}\\  \hline
\end{tabular}
\begin{tabular}{ccccccc}
\multicolumn{7}{c}{Feature}                                            \\  \hline
        dark  & light & natural & human & average & perfect & normal \\ \hline
 -0.02 & 0.00  & -0.01   & \textbf{0.02**}  & \textbf{0.02*}    & 0.00    & -0.01  \\
 0.23  & 0.27  & \textbf{1.76**}    & \textbf{-1.68**} & 0.29    & -0.39   & 1.03   \\
\textbf{1.67**}  & \textbf{1.40**}  & \textbf{2.76**}    & \textbf{1.05**}  & -0.01   & 0.21    & -0.04  \\
 -1.84 & -4.24 & -0.61   & \textbf{0.41**}  & -0.43   & -0.33   & \textbf{0.04*}   \\
 0.17  & -1.21 & -0.87   & \textbf{0.49**}  & -0.61   & -3.66   & -3.62  \\
 0.01  & 1.03  & 0.08    & 0.23  & 0.41    & 0.47    & -2.95  \\
\textbf{-6.76**} & \textbf{-7.73**} & \textbf{-8.05**}   & \textbf{-4.72**} & \textbf{-8.60**}   & \textbf{-8.21**}   & \textbf{-8.57**}      \\  \hline
\end{tabular}

\hfill \break \newline
\centering
\begin{tabular}{cccccccccc}
\multicolumn{8}{c}{Size}                                     &        \\  \hline
    &  big   & small & thin  & fat   & chubby & slim   & skinny & strong \\ \hline
age & \textbf{-0.01*} & \textbf{-0.04*} & -0.01 & 0.01  & \textbf{-0.07*}  & 0.00   & \textbf{0.07**}   & 0.03   \\
female  & 0.40  & 0.77  & \textbf{0.63**}  & -0.08 & -0.13  & 0.29   & 0.87   & -0.75  \\
black  & 0.11  & 0.90  & 0.05  & \textbf{1.01**}  & -0.02  & 4.27   & 2.08   & 1.55   \\
asian & -0.62 & 0.21  & -1.07 & 0.62  & -0.07  & 2.83   & -2.99  & 1.55   \\
indian  &  -0.09 & 1.14  & \textbf{-1.44**} & -0.01 & -3.40  & 3.88   & 2.38   & -2.98  \\
other  & -0.60 & 0.19  & -0.21 & 0.21  & -2.95  & -2.06  & -1.71  & -2.00  \\
$\beta_0$   &  \textbf{-6.26**} & \textbf{-7.97**} & \textbf{-6.11**} & \textbf{-7.29**} & \textbf{-6.91**}  & \textbf{-11.69**} & \textbf{-14.81**} & \textbf{-10.19**}     \\  \hline
\end{tabular}

\end{table}

\end{document}